\documentclass{aa}

\date{March 2022}
\usepackage{natbib}
\bibpunct{(}{)}{;}{a}{}{,} 
\usepackage{amssymb,amsmath}
\usepackage{mathtools}
\usepackage{bm}
\usepackage{soul}
\usepackage{txfonts}
\usepackage{graphicx}
\usepackage{color}
\definecolor{darkgreen}{rgb}{0,0.60,0}

\DeclarePairedDelimiterX\braket[2]{\langle}{\rangle}{#1 \delimsize\vert #2}
\begin{document}
\title{The polarization signals of the solar K~{\sc i} D lines \\ and their magnetic sensitivity}
\author{
E. Alsina Ballester\inst{1,2,3} 
}

\institute{
Instituto de Astrof\'{i}sica de Canarias, 38205 La Laguna, Tenerife, Spain
\and
Departamento de Astrof\'{i}sica, Universidad de La Laguna, 38206 La Laguna, Tenerife, Spain
\and
IRSOL Istituto Ricerche Solari ``Aldo e Cele Dacc\'{o}'', Universit\`{a} della Svizzera italiana, 6605 Locarno-Monti, Switzerland\\ 
\email{ernest.alsina@iac.es}
}           
 \abstract
{}
{This work aims to identify the relevant physical processes in shaping the intensity and polarization patterns of the solar K~{\sc{i}} D lines through spectral syntheses, placing particular emphasis on the D${}_2$ line.} 
 {The theoretical Stokes profiles were obtained by numerically solving the radiative transfer problem for polarized radiation considering one-dimensional semi-empirical models of the solar atmosphere. The calculations account for scattering polarization, partial frequency redistribution (PRD) effects, hyperfine structure (HFS), $J$- and $F$-state interference, multiple isotopes, and magnetic fields of arbitrary strength and orientation.} 
{The intensity and circular polarization profiles of both D lines can be suitably modeled while neglecting both $J$-state interference and HFS. 
The magnetograph formula can be applied to both lines, without including HFS, to estimate weak longitudinal magnetic fields in the lower chromosphere. By contrast, modeling the scattering polarization signal of the D lines requires the inclusion of HFS. 
The amplitude of the D${}_2$ scattering polarization signal is strongly depolarized by HFS, but it remains measurable. A small yet appreciable error is incurred in the scattering polarization profile if PRD effects are not taken into account. 
Collisions during scattering processes have a clear depolarizing effect, although a quantitative analysis is left for a forthcoming publication. 
Finally, the D${}_2$ scattering polarization signal is particularly sensitive to magnetic fields with strengths around $10$~G and it strongly depends on their orientation.  
Despite this, its center-to-limb variation relative to the amplitude at the limb is largely insensitive to the field strength and orientation. } 
{These findings highlight the value of the K~{\sc i} D${}_2$ line polarization for diagnostics of the solar magnetism, and show that the linear and circular polarization signals of this line are primarily sensitive to magnetic fields in the lower chromosphere and upper photosphere, respectively. }
 \date{}
 \keywords{Radiative transfer --
                Scattering --
                Polarization -- 
                Sun: chromosphere}
\maketitle
\section{Introduction}
 \label{sec::introduction} 
The K~{\sc{i}} D${}_1$ and D${}_2$ lines resonance lines encode valuable information on the layers of the solar atmosphere between the upper photosphere and the lower chromosphere, where the temperature minimum is found. 
The radiative transfer (RT) investigations on the K~{\sc{i}} D${}_1$ line carried out by \cite{Bruls+92}, without assuming local thermodynamic equilibrium (LTE), led to the conclusion that its line-core intensity profile is primarily formed in the photosphere. 
Furthermore, the height at which the line-center optical depth is unity was found at $450$~km above the solar surface, which is slightly below the temperature minimum \citep[see also the line-depression contribution functions in][]{BrulsRutten92}. 
A more recent theoretical investigation \citep[see][]{QuinteroNoda+17} on the two potassium D lines found that the temperature sensitivity of their  
line-core intensity is strongest at the lower photosphere. 
On the other hand, it was also found that the responses of the intensity to variations in the line-of-sight (LOS) velocity and of the circular polarization to the magnetic field were centered at heights corresponding to the upper photosphere and even the lower chromosphere. 
These differences were attributed to the fact that the temperature sensitivity comes from lower layers where the line source function is coupled to the local thermal contributions, whereas the sensitivity to the LOS velocity and the magnetic fields comes from higher layers where scattering processes play a more dominant role. 
In addition, the D${}_2$ line was found to be sensitive to variations in these properties at slightly greater atmospheric heights than the D${}_1$ line.  
The D lines of potassium thus appear to be excellent candidates for multiwavelength investigations; by observing them together with other spectral lines, we could potentially probe the properties of the solar atmosphere at various depths at the same time. 

In observations close to the limb, the D${}_2$ line is expected to show linear polarization signals of substantial amplitude, due to the scattering of anisotropic radiation, which induces population imbalances and coherence between the various magnetic sublevels of the atom (i.e., atomic level polarization). 
These so-called scattering polarization signals are known to be sensitive to the presence of magnetic fields via the Hanle effect \citep[e.g.,][]{TrujilloBueno01}, which increases the diagnostic value of the D${}_2$ line. 
Unfortunately, this line is strongly absorbed in the Earth's atmosphere by a molecular O${}_2$ line \citep[e.g., ][]{Delbouille+73}, and it is thus inaccessible to ground-based observations. 
However, steps have been made toward observing this line in recent years. One notable example is the SUNRISE Chromospheric Infrared spectroPolarimeter \citep[SCIP; see][]{Katsukawa+20}, designed to provide unprecedented spectro-polarimetric measurements of this line, among many others, during the flight of the SUNRISE III balloon-borne observatory. 

A number of RT investigations focused on the K~{\sc{i}} D lines were carried out in the past 
\citep{Bruls+92,BrulsRutten92,UitenbroekBruls92,QuinteroNoda+17}, although they did not take scattering polarization and its magnetic sensitivity into account. 
The hyperfine structure (HFS) of potassium was not considered either, even though its two most abundant stable isotopes, $^{39}\!$K and $^{41}\!$K, have nuclear spin $I = 3/2$ and even though all the atomic levels involved in the transitions responsible for the D${}_1$ and D${}_2$ lines 
present considerable hyperfine splitting \citep[e.g.,][]{Ney69,Bendali+81,NIST_ASD}. 
Although HFS is known to only have an appreciable impact on the intensity profiles for a small number of spectral lines, such as the infrared Mn~{\sc{i}} lines investigated by \cite{AsensioRamos+07}, it is essential for modeling the scattering polarization of various other resonance lines of interest, such as the Ba~{\sc{ii}} and Na~{\sc{i}} D lines \citep{Belluzzi+07,BelluzziTrujilloBueno13,AlsinaBallester+21}. 

\begin{figure}[!h]
 \centering
\includegraphics[width = 0.485\textwidth]{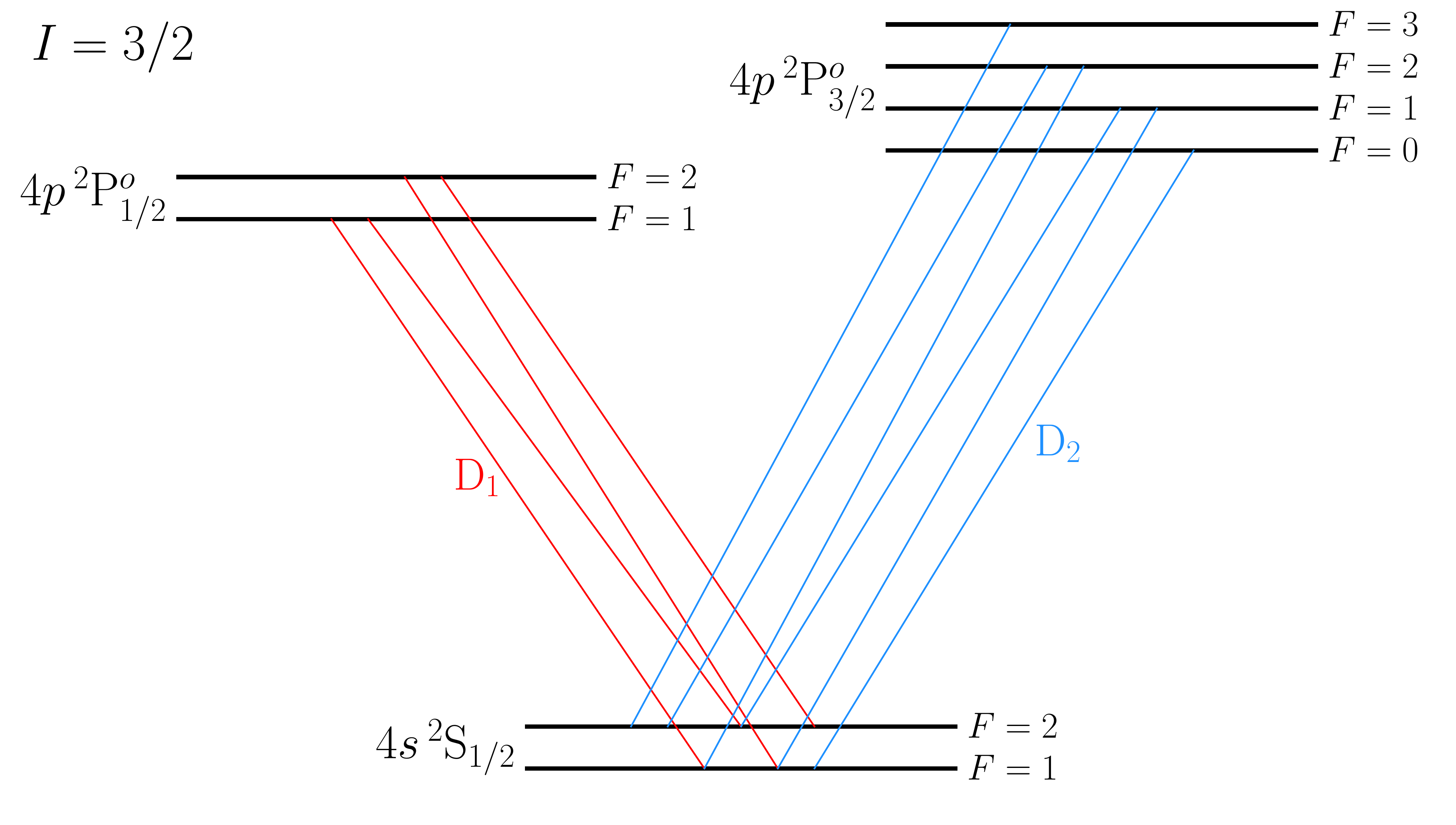}
\caption{Grotrian diagram for the two-term system with hyperfine structure (HFS) considered to model the K~{\sc{i}} D lines. The D${}_1$ and D${}_2$ lines arise from the transitions between states of the same lower fine structure (FS) level $4s \, {}^{2}\mathrm{S}_{1/2}$ (ground term) and the states of the $4p \, {}^{2}\mathrm{P}^{o}_{1/2}$ and $4p \, {}^{2}\mathrm{P}^{o}_{3/2}$ levels, respectively. The atomic system has nuclear spin $I = 3/2$, and thus the D${}_1$ line consists of four HFS components (red lines), whereas the D${}_2$ line consists of six HFS components (blue lines). } 
	\label{fig::Grotrian}%
\end{figure} 
As illustrated in Fig.~\ref{fig::Grotrian}, the K~{\sc{i}} D lines arise from the radiative transitions between the states of the ground term of potassium, $4s \, {}^{2}\mathrm{S}$, and the upper term $4p \, {}^{2}\mathrm{P}^{o}$. The ground term consists of a single fine structure (FS) level with $J = 1/2$, which is common to the two lines. The upper term consists of two FS levels with $J = 1/2$ and $J = 3/2$, which are the upper levels of the D${}_1$ and D${}_2$ lines, respectively.
When accounting for the nuclear spin, these FS levels are further split into various HFS levels characterized by quantum number $F$. 
Thus, the D${}_1$ and D${}_2$ lines each consist of several HFS transitions. 
The main goal of the present article is to investigate which physical mechanisms are relevant in shaping the intensity and polarization profiles of these lines. 
For this purpose, the spectral synthesis code introduced in \cite{AlsinaBallester+22}, hereafter ABT22, was used. This code relies on the numerical solution of the transfer problem for polarized radiation under non-LTE conditions. 
It is based on the theoretical framework of \cite{Bommier17} and is suitable for modeling spectral lines that arise from a two-term system that includes HFS. 
Because the code considers a two-term model, the quantum interference between FS or HFS levels that belong to the same term can be taken into account. 
Partial frequency redistribution (PRD) effects, or the frequency correlations between incoming and outgoing radiation in scattering processes, are also accounted for. 
In addition, it takes into account the influence of magnetic fields of arbitrary strengths and orientations considering the Paschen-Back (PB) effect.  

The article is structured as follows. Section~\ref{sec::formulation} presents the main equations involved in the considered RT problem and a brief discussion on the numerical methods employed in the above-mentioned numerical code. Section~\ref{sec::nonmag} is dedicated to the evaluation of the impact of specific (nonmagnetic) physical parameters or processes on the intensity and polarization patterns of the K~{\sc{i}} D lines. First, the impact of accounting for the second-most abundant of the stable isotopes, $^{41}\!$K, is examined. Then, the influence of considering atmospheric models that represent regions of the Sun with different levels of activity is explored. 
The redistribution and depolarization effects of collisions with neutral perturbers are also studied in this section. Finally, the impact on the line profiles of HFS and the quantum interference between different atomic states is investigated. The analysis in Sect.~\ref{sec::magnetic} is focused on the impact of the magnetic field, first regarding the scattering polarization signals of the D${}_2$ line in the presence of magnetic fields of various geometries, considering an LOS close to the solar limb. 
The same section also considers an analysis of the circular polarization patterns in both D lines, highlighting which physics mechanisms must be taken into account to suitably model them. 
Section~\ref{sec::CLV} presents a brief analysis of the center-to-limb (CLV) variation of the intensity and linear polarization at the center of the D${}_2$ line. 
In Sect.~\ref{sec::AppResponse}, the response functions for the intensity and polarization of the K~{\sc{i}} D${}_2$ line are presented and discussed when considering atmospheric models permeated with vertical and horizontal magnetic fields, accounting for the first time for scattering polarization and HFS. Finally, the main conclusions of this work are summarized in Sect.~\ref{sec::conclusions}. 

The parameters of the atomic model considered in the RT calculations, and details of the numerical calculations, are given in Appendix~\ref{sec::AppAtomicProps}. The extension to the case of multiple isotopes of the expressions that are implemented in the RT code, which are given in ABT22, is detailed in Appendix~\ref{sec::AppIsotopes}. 
The Hanle diagram for the scattering polarization emitted at a height in the atmospheric model that corresponds to the lower chromosphere, and its variation with the strength and orientation of the magnetic field is shown in Appendix~\ref{sec::AppHanleDiag}. The Stokes profiles for the D lines, calculated for the same geometries considered in Sect.~\ref{sec::magnetic} but in the presence of stronger magnetic fields, are shown in Appendix~\ref{sec::AppExtraFig}.  

\section{Formulation of the problem}
 \label{sec::formulation}
The spectral synthesis code used throughout this work solves the non-LTE RT problem in one-dimensional (1D) atmospheric models. 
The numerical framework on which the code is based is described in detail in ABT22. 
For the sake of brevity this section only contains the expressions for the main quantities involved in the solution of the RT problem and a brief discussion on the considered iterative scheme and on the geometry of the problem. 
 \subsection{{Equations of the problem}}  
\label{sec::BasEq} 
The intensity and polarization of a partially polarized radiation beam is characterized by its Stokes vector $\boldsymbol{I}$, whose components $I_i$ ($i=\{0,1,2,3\}$) correspond to the Stokes parameters $I$, $Q$, $U$, and $V$, respectively. 
The variation of the Stokes vector at frequency $\nu$ as it propagates through the medium in direction $\mathbf{\Omega}$ along the ray path, determined by coordinate $s$, is given by the radiative transfer equation (RTE) 
 \begin{equation}
  \frac{\mathrm{d}}{\mathrm{d}s} \boldsymbol{I}(\nu, \mathbf{\Omega}) = - \mathbf{K}(\nu, \mathbf{\Omega}) \, \boldsymbol{I}(\nu, \mathbf{\Omega}) 
  +\boldsymbol{\varepsilon}(\nu, \mathbf{\Omega})  . 
  \label{eq::RTE}
 \end{equation}
All the quantities that appear in this ordinary differential equation have a dependency on the spatial coordinate 
$\mathbf{r}$,\footnote{{The spatial coordinates $\mathbf{r}$ and $s$ are related through 
$\mathrm{d}s = \mathbf{\Omega} \cdot \mathrm{d} \mathbf{r}$.}} which is not shown explicitly in this article for  simplicity of notation. 
The propagation matrix $\mathbf{K}$ has the form  
 \begin{equation}
 \mathbf{K} = \left(
 \begin{array}{c c c c}
  \eta_I & \eta_Q & \eta_U  & \eta_V \\
  \eta_Q & \eta_I & \rho_V  & -\rho_U \\
  \eta_U & -\rho_V & \eta_I & \rho_Q \\
  \eta_V & \rho_U & -\rho_Q & \eta_I
 \end{array} \right) \, , 
 \label{eq::Prop}
\end{equation}
in which the dependencies on $\nu$ and $\mathbf{\Omega}$ are not shown for simplicity of notation. 
The diagonal element $\eta_I$ is the absorption coefficient. 
The $\eta_Q$, $\eta_U$, and $\eta_V$ elements are the dichroism coefficients, and $\rho_Q$, $\rho_U$, and $\rho_V$ are the anomalous dispersion coefficients.  
The emission in each of the four Stokes parameters is quantified by the four components of the emission vector $\boldsymbol{\varepsilon}$. 

Hereafter, the elements of the propagation matrix and the components of the emission vector are collectively referred to as the RT coefficients. In general, each of them has a line and a continuum contribution, but under typical conditions of the solar atmosphere the continuum contribution to the dichroism and anomalous dispersion coefficients can be safely neglected \citep[see][hereafter LL04]{BLandiLandolfi04}. 
Thus, the absorption coefficient and the $i$-th component of the emission vector can be decomposed into 
\begin{align}
  \eta_I(\nu,\mathbf{\Omega}) = \eta^\ell_I(\nu,\mathbf{\Omega}) + \eta^c_I(\nu,\mathbf{\Omega}) \, ,
 \label{eq::EtaLineCont} \\ 
 \varepsilon_i(\nu,\mathbf{\Omega}) = \varepsilon_i^{\ell}(\nu, \mathbf{\Omega}) + \varepsilon_i^{c}(\nu, \mathbf{\Omega}) \, , 
 \label{eq::EmisLineCont}    
\end{align}
where the labels $\ell$ and $c$ indicate the line and continuum contribution, respectively. Each of these contributions to the emission vector can be further decomposed into scattering and thermal contributions, labeled ``$\mbox{sc}$'' and ``$\mbox{th}$,'' respectively. The continuum contributions considered in this work are obtained as explained in ABT22. 

Regarding the line contribution, it is worth recalling that the atomic system considered in this work is, in the most general case, a two-term atom with HFS.\footnote{Unlike in the case of a multilevel atomic model, the quantum interference between FS levels that belong to the same term is taken into account in a multiterm system.} In the absence of a magnetic field, each of the atomic states of the system can be characterized by the quantum numbers for the electronic (orbital plus spin) total angular momentum $J$ and the atomic (electronic plus nuclear) total angular momentum $F$. The various (degenerate) states with the same numbers $J$ and $F$ are further characterized by the magnetic quantum number $f$, which corresponds to a measurement of the operator for $F$ along a selected quantization axis. In the nonmagnetic case, one can make a distinction between the quantum interference between pairs of states that belong to (i) the same $J$ and $F$ level, (ii) different $F$ levels of the same $J$ level -- hereafter $F$-state interference -- and (iii) different $J$ levels -- hereafter $J$-state interference \citep[e.g.,][]{Stenflo97,CasiniMansoSainz05}. As will be discussed further in Sect.~\ref{sec::magnetic}, the degeneracy of the $f$ states is broken in the presence of magnetic fields. When the magnetic field is strong enough that the incomplete PB effect regime is reached, $J$ and $F$ are no longer good quantum numbers of the system and thus they cannot be used to characterize the atomic states. 

In addition, the lower term of the system is assumed to be unpolarized and infinitely sharp.\footnote{An atomic term is said to be polarized when the various states that belong to it present population imbalances or quantum coherence. The term is said to be infinitely sharp if all its states are infinitely sharp.} 
Such an atomic system is generally suitable for modeling resonance lines that originate from transitions between the ground term of the atom and an upper term that is not collisionally or radiatively connected to other terms of lower energy, which is the case for the K~{\sc{i}} D lines. Furthermore, the abovementioned code can account for the contribution from different isotopes.  
The expressions for the line contributions to the elements of the propagation matrix, for an atomic system with a single isotope, can be found in Appendix~C.3 of ABT22. The generalization to a two-isotope atomic model is straightforward (see Appendix~\ref{sec::AppIsotopes}).  

In LTE conditions, Kirchhoff's law states that the unpolarized line emissivity divided by the absorption coefficient is equal to the Planck function. 
According to the principle of detailed balance, this relation must hold separately for any direction, frequency range, and polarization state \citep[e.g., Sect.~9.15 in][]{BReif65}. Thus, in LTE conditions, the polarized components of the line emission vector are proportional to the corresponding dichroism coefficients. Moreover, these relations for the emissivity can be separated into the scattering and thermal contributions, 
and the latter is given by 
\begin{equation}
\varepsilon_i^{\ell, \mbox{\scriptsize th}}(\nu,\mathbf{\Omega}) = \epsilon \, B_T(\nu) \, \eta_i(\nu,\mathbf{\Omega}) \, ,
\label{eq::EmisLineTherm}
\end{equation}
where $B_T(\nu)$ is the Planck function, taking the Wien limit (under the reasonable assumption of neglecting stimulated emission). The photon destruction probability due to inelastic collisions is given by $\epsilon = \Gamma_I/(\Gamma_R + \Gamma_I)$, where $\Gamma_R$ and $\Gamma_I$ are the contributions to the line broadening constant from radiative transitions and inelastic collisions, respectively. This expression is independent of the radiation field and holds in non-LTE conditions as long as inelastic collisions are isotropic and the perturbers follow a Maxwellian distribution of velocities. The expression for the particular case of a two-level atom was derived in Sect.~10.6 of LL04 and the one for a multiterm atom with HFS in the nonmagnetic case can be found in \cite{Belluzzi+15}. The corresponding derivation for a two-term atom with HFS accounting for the PB effect will be presented in a separate paper. 

The line scattering contribution to the emission vector can be related directly to the incident radiation, making use of the redistribution matrix formalism \citep[e.g.,][]{DomkeHubeny88} 
\begin{equation}
  \varepsilon_i^{\ell, \mbox{\scriptsize sc}}(\nu,\mathbf{\Omega}) = \, k_M \!\int\!\!\mathrm{d}\nu^\prime \! \oint \mathrm{d}\mathbf{\Omega}^\prime \!
 \sum_{j = 0}^3 \bigl[\mathcal{R}(\nu^\prime,\mathbf{\Omega}^\prime, \nu, \mathbf{\Omega}) \bigr]_{i j} \, I_j(\nu^\prime,\mathbf{\Omega}^\prime) \, ,    
 \label{eq::EmisLineScat}
\end{equation}
where the primed quantities correspond to the incident radiation. The labels $i$ and $j$ indicate the Stokes components of the scattered and incident radiation, respectively. 
The quantity $k_M$ is the frequency-integrated absorption coefficient at the considered spatial point $\mathbf{r}$ (e.g., Sect.~10.16 of LL04). 
An expression for the redistribution matrix can be obtained for the considered atomic system, because a closed analytical solution exists for the master equation that characterizes the populations of its various states and the quantum interference between pairs of them. 

As explained in detail in ABT22, the RT code used for this investigation makes use of the redistribution matrices presented in \cite{Bommier17,Bommier18}. Following the notation introduced in \cite{Hummer62}, the redistribution matrix can be given as a sum of the $\bm{\mathcal{R}_{\mbox{\sc{ii}}}}$ and $\bm{\mathcal{R}_{\mbox{\sc{iii}}}}$ matrices. 
The $\bm{\mathcal{R}_{\mbox{\sc{ii}}}}$ matrix quantifies the scattering processes for which frequency coherence is preserved in the atomic rest frame,  and $\bm{\mathcal{R}_{\mbox{\sc{iii}}}}$ quantifies those for which the frequencies of the incoming and outgoing radiation are uncorrelated in the same reference frame, due to collisions that perturb the atomic system. 
The RT calculations are carried out in the reference frame of the observer, taking into account the frequency redistribution effects produced by small-scale atomic motions via the Doppler effect. This introduces an angle-frequency coupling which greatly increases the numerical cost of computing the emission vector. %
In order to avoid this coupling, the following approximations are used. For $\bm{\mathcal{R}_{\mbox{\sc{ii}}}}$, the angle-averaged approximation introduced in \cite{ReesSaliba82} is applied. 
The assumption that the frequencies of the incident and scattered radiation are completely uncorrelated in the observer's reference frame is made for $\bm{\mathcal{R}_{\mbox{\sc{iii}}}}$; this is often referred to as the assumption of complete frequency redistribution (CRD) in the observer's frame. Under these approximations, the expressions for the redistribution matrices for the above-mentioned atomic system are found in Appendix~C.4 of ABT22, which can also be easily generalized to the two-isotope case (see Appendix~{\ref{sec::AppIsotopes}}). 

Appreciable errors in numerical calculations due to the angle-averaged approximation in $\mathcal{R}_{\mbox{\sc{ii}}}$ were reported in the past, both when considering semi-infinite and isothermal atmospheric models \citep{Faurobert88,Sampoorna+17} and when synthesizing strong resonance lines in semi-empirical atmospheric models \citep[e.g.,][]{Janett+21}.  
The suitability of assuming CRD in the observer's reference frame for $\bm{\mathcal R}_{\mbox{\sc{iii}}}$ in the polarized case was called into question by \cite{Bommier97b}, but more recent RT investigations by \cite{Sampoorna+17} suggest that that the incurred error is minor when considering atmospheres with a large optical depth relative to the considered line. However, a quantitative determination of the accuracy of the approximations for $\bm{\mathcal{R}}_{\mbox{\sc{ii}}}$ and $\bm{\mathcal{R}}_{\mbox{\sc{iii}}}$ is beyond the scope of this investigation. 
These approximations are suitable for the goal of this paper, which is to determine which physical mechanisms shape the intensity and polarization profiles of the K~{\sc{i}} D lines, rather than to accurately reproduce observations. 
 
 \subsection{The numerical scheme}
 \label{sec::NumScheme}
The theoretical line profiles shown in the following section are obtained from the numerical solution of the 
non-LTE RT problem for polarized radiation. 
The employed synthesis code iteratively solves the RTE (Eq.~\eqref{eq::RTE}) and the equation that relates the radiation field to the line emissivity through the redistribution matrices (Eq.~\eqref{eq::EmisLineScat}) until a self-consistent solution is reached. As explained in Sect.~2.6 of ABT22, the problem is solved through a two-step process that consists of two distinct RT calculations. 

The goal of the first step is to provide the population of the lower term $N_\ell$, which is kept fixed in the second step, where it is used to compute $k_M$. 
{In this work, this preliminary step was carried out in the unpolarized case and considering a multilevel atom (see Appendix~\ref{sec::AppAtomicProps}), making use of the RH code of \cite{Uitenbroek01}}. This calculation provides a more realistic value of $N_\ell$ than the one that would be obtained by considering a two-term atomic model. 
This step also provides other quantities of interest including the rates of elastic and inelastic collisions, the continuum absorption coefficient $\eta^c_I$, the unpolarized continuum thermal emissivity $\varepsilon^{c, \mbox{\scriptsize th}}_I$, and the extinction coefficient for the scattering of the continuum $\sigma_c$. These quantities are stored and used in step 2 of the calculation. 
 
In the second (or main) step, the RT problem is solved while taking polarization into account, considering in the most general case a two-term atomic model with HFS and the two most abundant stable isotopes of potassium. 
The step-2 RT calculations are carried out using the code described in Sect.~2 of ABT22, which makes use of expressions shown in the previous subsection. 
These calculations are carried out using the Jacobi iterative scheme described in the same paper, and Eq.~\eqref{eq::RTE} is solved numerically via the DELOPAR short-characteristics method \citep[see][]{TrujilloBueno03}.  

\begin{figure*}[!t]
\centering
 \includegraphics[width = 0.975\textwidth]{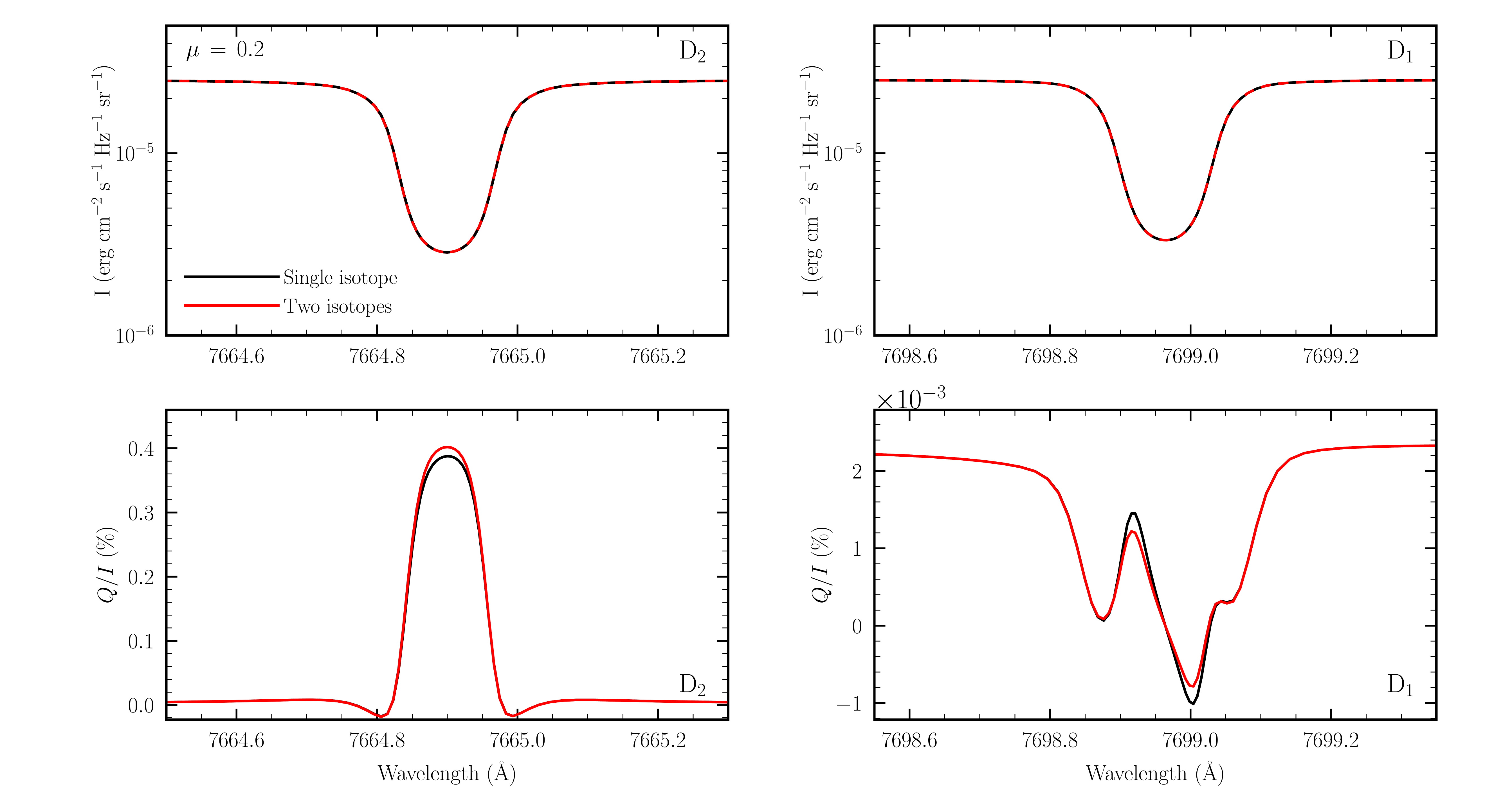}
  \caption{Intensity (upper panels) and Stokes $Q/I$ (lower panels) profiles as a function of air wavelength, for the $0.8$ \AA--wide spectral ranges centered on the K~{\sc i} D${}_2$ (left panels) and D${}_1$ (right panels) lines. Calculations were carried out using the radiative transfer (RT) code introduced in Sect.~\ref{sec::formulation}, considering a two-term atomic model with HFS, accounting for partial frequency redistribution (PRD) effects, and in the absence of a magnetic field. For all the figures in this work, a line of sight (LOS) with $\mu = 0.2$ is taken except where otherwise noted. 
The results of calculations considering the two most abundant isotopes of potassium (solid red curves) are compared to those in which only $^{39}$\!K was considered (solid black curves). The atmospheric model C of \cite{Fontenla+93}, hereafter FAL-C, was considered for both calculations. 
  Throughout this work, the reference direction for positive Stokes $Q$ is taken to be parallel to the limb.} 
  \label{fig::isotopes} 
\end{figure*}

\subsection{Atmospheric model and geometry of the problem}
In all calculations, 1D semi-empirical models of the solar atmosphere were considered. 
Except where otherwise noted, model C of \cite{Fontenla+93}, hereafter FAL-C, was considered; this model is representative of a spatially averaged region of the quiet Sun. 
The LOS of the emergent radiation is given relative to the local vertical, through $\mu = \cos\theta$, in which $\theta$ is the inclination or heliocentric angle. Except where otherwise noted, an LOS with $\mu = 0.2$ was considered; this corresponds to radiation emerging close to the solar limb. 
Magnetic fields, when included (see Sect.~\ref{sec::magnetic}), were taken with a fixed strength and orientation at all height points of the model. Their orientation is determined by the inclination relative to the local vertical, $\theta_B$, and the azimuth $\chi_B$, or the angle relative to the plane defined by the local vertical and the LOS. 
The reference direction for positive Stokes $Q$ was taken to be parallel to the limb except where otherwise stated (see Fig.~1 of ABT22). 

\section{The intensity and polarization of the K~{\sc i} D lines in the nonmagnetic case} 
\label{sec::nonmag}
{This section is dedicated to analyzing which physical mechanisms and properties play an important role in the intensity and polarization profiles of the K~{\sc{i}} D lines, especially D${}_2$, in the absence of magnetic fields. In order to identify the impact of a specific physical mechanism, the results of calculations in which it was taken into account and neglected are compared.  
In the most general calculations, the two most abundant stable isotopes of potassium, HFS, both $J$- and $F$-state interference, and PRD effects in scattering were all taken into account. 
The resulting intensity and $Q/I$ profiles are represented by the red curves in Fig.~\ref{fig::isotopes}. These profiles are shown as a function of air wavelength for two spectral ranges centered on the D${}_1$ and D${}_2$ lines, each with a width of $0.8$~\AA . The intensity signal at the center of the D-lines presents a drop of roughly one order of magnitude relative to the continuum value, and it is slightly deeper for the stronger D${}_2$ line. 
For the considered LOS, a $Q/I$ signal with an amplitude of roughly $0.4\%$ is found in the core of the D${}_2$ line. 
This scattering polarization signal arises from the anisotropic radiation that illuminates the atom, which induces atomic alignment.\footnote{Atomic alignment is a type of atomic level polarization and it refers to the population imbalances between states with different magnetic quantum numbers $\bigl|f\bigr|$ that belong to the same atomic level. 
Atomic orientation refers to the population imbalances between the states with quantum numbers $f$ and $-f$.} 
The antisymmetric $Q/I$ pattern found within the core of the D${}_1$ line arises from a related but distinct physical mechanism. 
It originates because the incident anisotropic radiation field is spectrally structured over the range spanned by the different HFS transitions that contribute to the D${}_1$ line \citep[see][]{BelluzziTrujilloBueno13,AlsinaBallester+21}. 
However, because the hyperfine splitting in both the upper and lower levels of the D${}_1$ line is relatively small, the maximum fractional linear polarization amplitude of the signal within the line core is on the order of $10^{-5}$, which is at the detection limit of even the latest generation of spectro-polarimeters \citep{Ramelli+10}. 
Thus, unlike the linear polarization pattern of the Na~{\sc{i}} D${}_1$ line \citep[e.g.,][]{AlsinaBallester+21}, the scattering polarization signal of the K~{\sc{i}} D${}_1$ line is too small to be of present diagnostic interest and thus it will be omitted from most of the discussions in the rest of this paper. 

\subsection{The influence of multiple isotopes on the modeling}
\label{sec::Isotopes}
Potassium is mostly found in the form of two stable isotopes, $^{39}$K and $^{41}$K, which have isotopic abundances of $93.3\%$ and $6.7\%$, respectively \cite[see][]{NIST_ASD}. When considering a single isotope instead of both of them, the numerical cost of computing the redistribution matrix decreases by roughly a factor 2. 
A comparison between the results of the RT calculations accounting for both isotopes (red curves) and only for $^{39}$K {by} setting its abundance to $100\%$ (black curves) is shown in Fig.~\ref{fig::isotopes}. 

\begin{figure*}[!t]
\centering
\includegraphics[width = 0.975\textwidth]{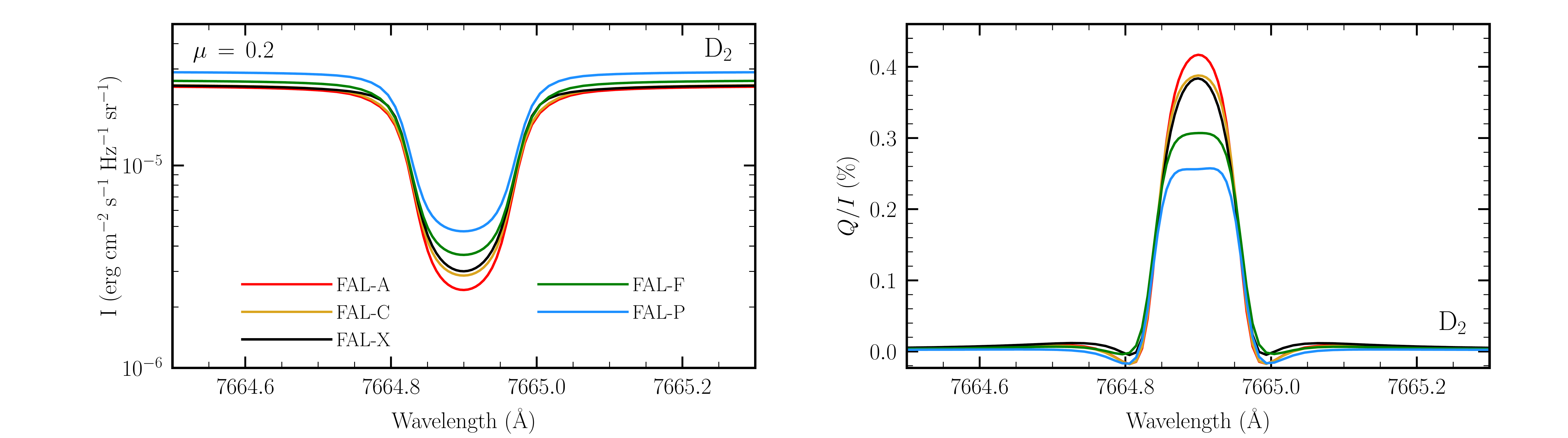}
\caption{Intensity (left panel) and Stokes $Q/I$ (right panel) profiles as a function of wavelength in the spectral range centered on the K~{\sc i} D${}_2$ line. The RT calculations were carried out in the absence of a magnetic field for a two-term atomic model with HFS, taking PRD effects into account.
In the calculations illustrated in this figure and all those shown below, only the ${}^{39}$K isotope was considered. 
The colored curves (see legend) represent the results of calculations considering the various atmospheric models presented in \cite{Fontenla+93} and \cite{Avrett95}. } 
  \label{fig::FAL} 
\end{figure*}
\begin{figure*}[!t]
\centering 
 \includegraphics[width = 0.975\textwidth]{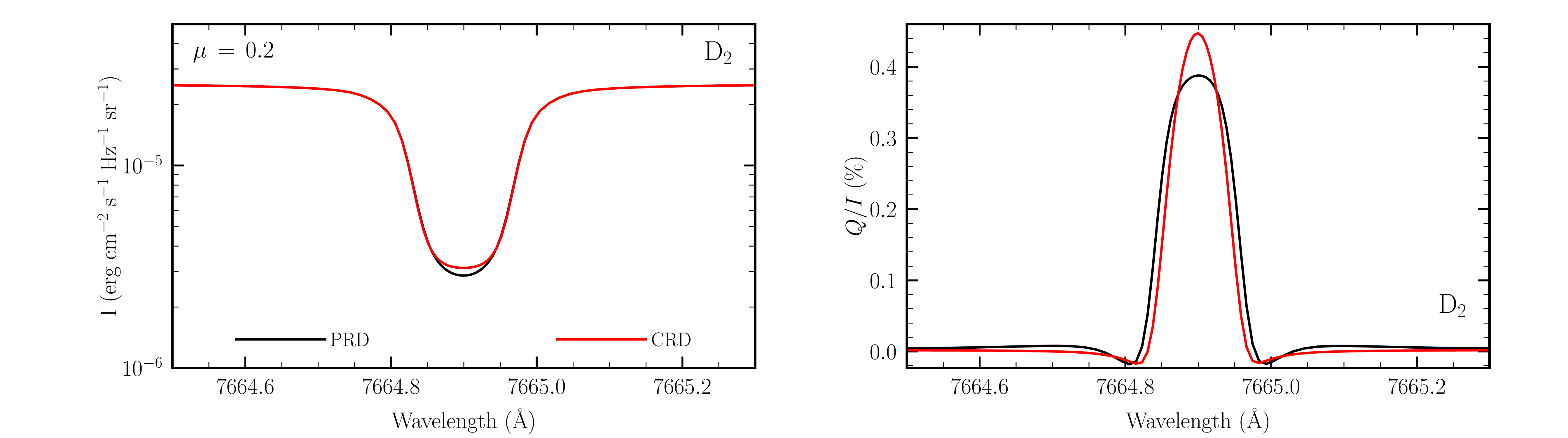}
  \caption{Intensity (left panel) and Stokes $Q/I$ (right panel) profiles as a function of wavelength in the spectral ranges centered on the K~{\sc i} D${}_2$ line. The RT calculations were carried out in the absence of a magnetic field and for a two-term atomic model with HFS. The results of the calculation accounting for PRD (black curves) are compared to those obtained in the complete frequency redistribution (CRD) limit (red curves) as described in the text.  In this figure and all those shown below, the calculations were carried out considering the FAL-C atmospheric model. }
  \label{fig::CRDCS} 
\end{figure*}
No differences are found between the intensity profiles obtained from the two calculations. 
On the other hand, accounting for the ${}^{41}$K isotope has a small but appreciable impact on the linear polarization signal of D lines. At the center of the D${}_2$ line, the amplitude of the $Q/I$ signal is slightly above $0.40\%$ when it is included, and drops to slightly below $0.39\%$ when only considering ${}^{39}$K. A similar relative difference is found in the core of the D${}_{1}$ line. 
These variations can be taken to be acceptably small, and thus only the $^{39}$K isotope will be considered for the calculations carried out in the rest of the article. 
Moreover, the intensity profiles will be omitted from the rest of the figures if they coincide with those represented by the black curves in Fig.~\ref{fig::isotopes}.  

\subsection{The influence of the atmospheric models}
\label{sec::AtMod}
This section is dedicated to studying how the intensity and linear polarization signals of the K~{\sc{i}} D lines vary according to the level of activity of the considered region of the solar atmosphere. 
Figure~\ref{fig::FAL} shows a comparison between the $I$ and $Q/I$ profiles of the D${}_2$ line obtained from RT calculations for various semi-empirical 1D atmospheric models in the absence of magnetic fields. The corresponding profiles for the D${}_1$ line are not shown because the behavior of the $I$ profiles is very similar to the ones for D${}_2$ and, as noted above, the amplitude of its $Q/I$ is too small to be of present diagnostic interest. The considered models are A, C, F, and P of \cite{Fontenla+93}, hereafter the FAL models, which are representative of various regions of the solar atmosphere, ordered from the lowest to the highest level of activity. The M${}_{\mbox{\scriptsize CO}}$ model of \cite{Avrett95}, hereafter FAL-X, which represents a quiet region of the solar atmosphere with lower temperatures than FAL-C, was also considered. As expected, the intensity increases strongly with the level of activity, both in the continuum and the core of either lines. On the other hand, the drop in the line center relative to the continuum is smaller for models that correspond to more active regions. 

The amplitude of the $Q/I$ signal in the core of the D${}_2$ line decreases when considering models that correspond to regions with a higher level of activity. This amplitude is directly related to the degree of anisotropy of the radiation field at the atmospheric layers where the line forms. These profiles become less peaked as the level of activity increases; for the FAL-P model, a plateau is found around the line center. 

\subsection{The {impact} of frequency redistribution}
\label{sec::redis} 
When modeling the scattering polarization profiles of resonance lines that are stronger than the potassium D lines, such as the Ca~{\sc{i}} line at $4227$~\AA\ or the Mg~{\sc{ii}} doublet, it is crucial to account for PRD effects in order to reproduce the broad and large-amplitude signals produced in their wings \citep[e.g.,][]{Auer+80,BelluzziTrujilloBueno12,AlsinaBallester+18}.  
On the other hand, the approximation of neglecting frequency correlations between incoming and scattered radiation (i.e., CRD) is instead found to accurately reproduce the intensity and $Q/I$ signals of lines weaker than K~{\sc{i}} D${}_1$ and D${}_2$ 
that do not present broad scattering polarization wings, such as the photospheric Sr~{\sc{i}} line at $4607$~\AA\ \citep[e.g.,][]{AlsinaBallester+17}. 

This subsection analyzes the impact of PRD effects on the K~{\sc{i}} D${}_2$ line by comparing the theoretical profiles obtained by fully accounting for such effects, as in Sect.~\ref{sec::Isotopes}, and by making the CRD approximation. This approximation was implemented in the RT calculations by modifying the line broadening coefficient due to elastic collisions $\Gamma_E$ that enters the branching ratios of the redistribution matrices so that it is much larger than the radiative line broadening and thus practically all line scattering processes are quantified by $\bm{\mathcal{R}}_{\mbox{\sc{iii}}}$ (see Appendix~C.5 of ABT22). The resulting intensity and $Q/I$ profiles of the D${}_2$ line are shown in Fig.~\ref{fig::CRDCS}. 
The intensity profiles obtained through the two calculations are found to be in very good agreement, which is consistent with  \cite{UitenbroekBruls92} and \cite{QuinteroNoda+17}. 
The only notable difference is that the drop in the line-center intensity is slightly less pronounced for the CRD calculation than for the PRD one, which is common for absorption lines \citep[e.g.,][]{AlsinaBallester+17}. 
The rates of elastic collisions are relatively low where the D${}_2$ line-center optical depth is close to unity (between roughly $500$ and $650$~km in the FAL-C model, depending on the considered LOS) and most scattering processes are described by $\bm{\mathcal R}_{\mbox{\sc{ii}}}$ {(the coherence fraction, defined as in Sect.~10.4 of \citealt{BHubenyMihalas15}, ranges between $0.6$ and $0.8$ at the same heights in FAL-C)}. 
In spite of this, the difference between the PRD intensity profile and the one obtained under the CRD approximation 
remains very minor \citep[see also][]{Faurobert87}; their similarity can be attributed to the effects of Doppler redistribution {\citep[e.g.,][]{Thomas57}}. Although it is not shown here, the impact of CRD on the D${}_1$ intensity profile is likewise minor. 

On the other hand, the treatment of frequency redistribution has an appreciable impact in the linear polarization profile of the D${}_2$ line. 
In the line core, clear discrepancies are found between the $Q/I$ signals obtained from the two calculations. 
The maximum amplitude reaches $\sim\!0.45\%$ under the CRD approximation and $\sim\!0.39\%$ when accounting for PRD effects. 
At the D${}_2$ line-core formation depths, the anisotropy of the radiation field is smaller in the core than in the wings. This can explain why the line-core $Q/I$ is slightly smaller when accounting for frequency correlations than in the CRD case, in which only the wavelength-averaged radiation field is considered. 
Although it is not shown here, it can be verified that the differences between the $Q/I$ profiles obtained under the PRD and CRD treatments would remain appreciable when accounting for the smearing due to the resolution of a typical instrument \citep[for instance, $R = 2 \cdot 10^{5}$ for SCIP; see][]{Katsukawa+20}. 
Thus, in the calculations presented in the following sections, PRD is considered except where otherwise noted. 
\begin{figure}[!t]
 \centering
\includegraphics[width = 0.485\textwidth]{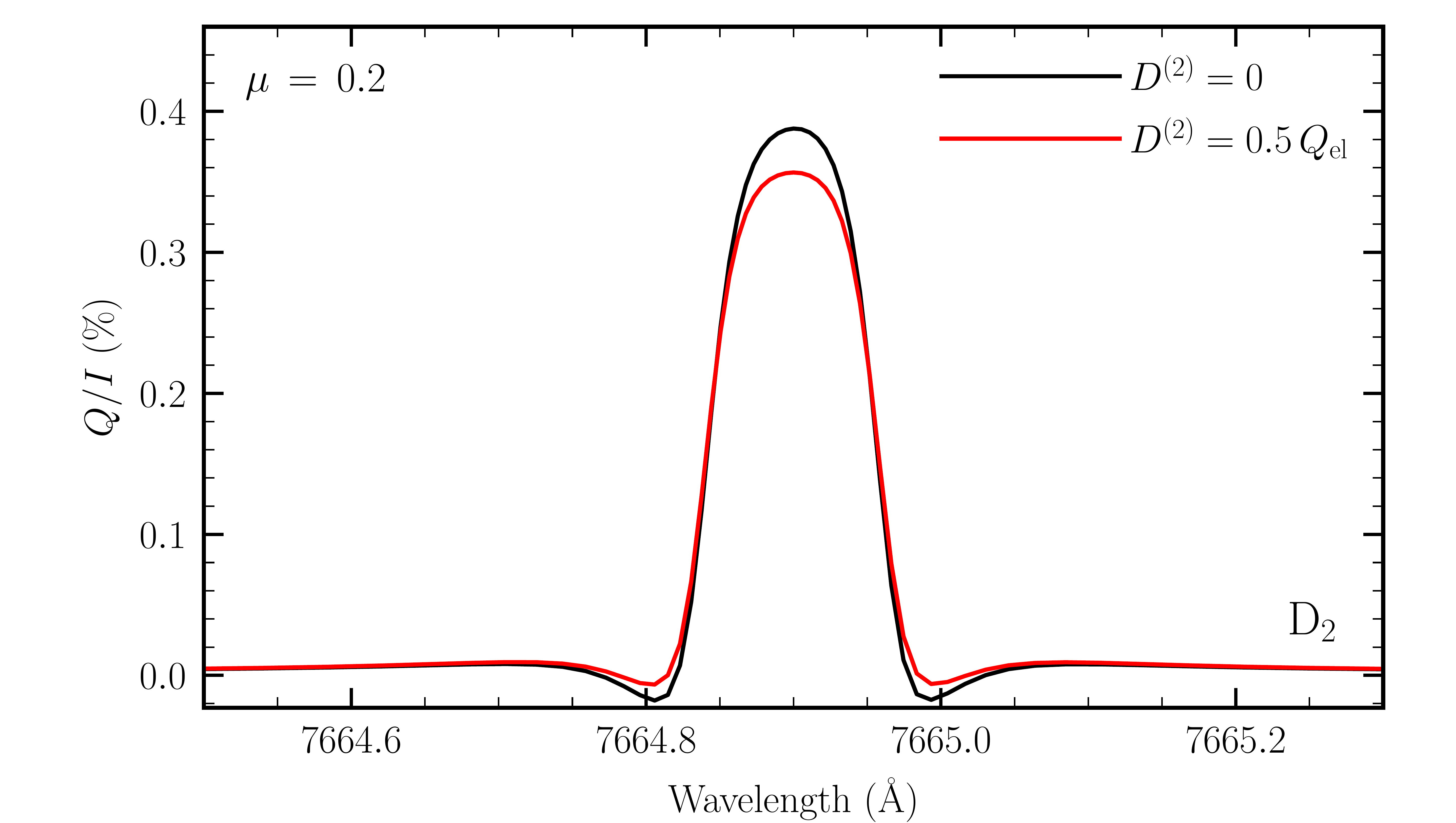}
\caption{Stokes $Q/I$ profile as a function of wavelength in the spectral range centered on the K~{\sc i} D${}_2$ line. The RT calculations were carried out in the absence of a magnetic field and considering a two-term atomic model with HFS. For the figures shown in the rest of this work, PRD effects were taken into account in the line scattering emissivity unless otherwise noted. Calculations carried out neglecting the impact of depolarizing collisions (black curves) are compared to those in which the $D^{(2)}(J_u,F_u)$ rates were set to $\frac{1}{2} Q_{\mathrm{el}}$ for all HFS levels with quantum number $F \ge 1$ (red curves). }
	\label{fig::collisions}%
\end{figure} 

\begin{figure*}[!t]
\centering
\includegraphics[width = 0.975\textwidth]{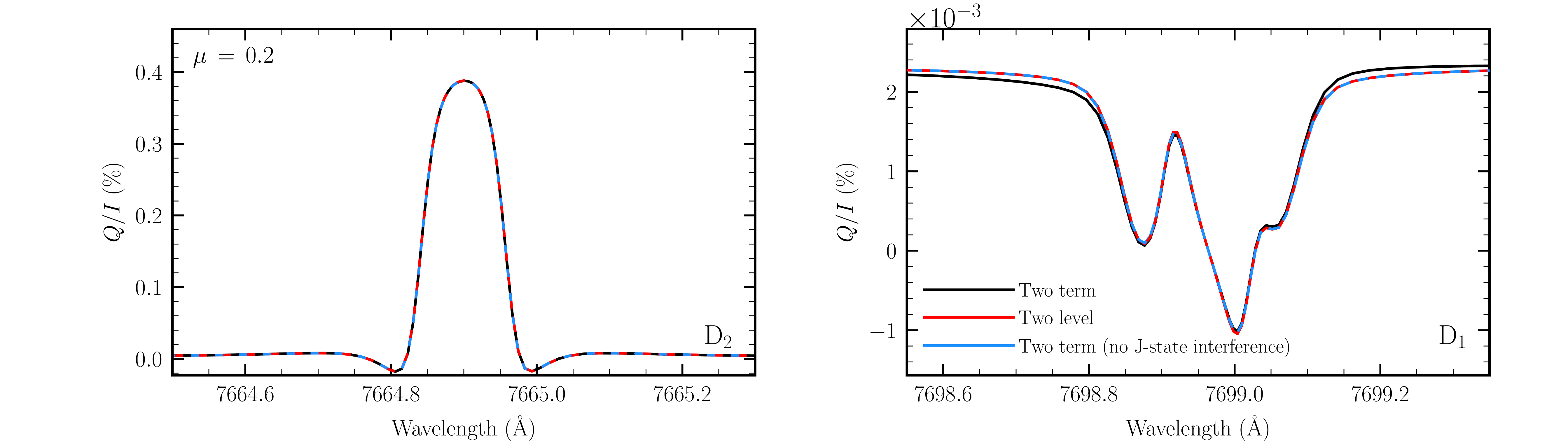}
\caption{Stokes $Q/I$ profiles as a function of wavelength in the spectral ranges centered on the K~{\sc i} D${}_2$ (left panel) and D$_{1}$ (right panel) lines. The RT calculations were carried out in the absence of a magnetic field. A comparison is shown between the calculations considering a two-term atomic model with HFS, both taking $J$-state interference into account (black curves) and neglecting it (blue curves), and the calculations considering two-level atomic models with HFS (red curves), in which there is no $J$-state interference by definition. Overlapping curves are dashed to enhance visibility.} 
  \label{fig::finestruc} 
\end{figure*}
\subsection{The depolarizing effect of elastic collisions} 
In the formation regions of the K~{\sc{i}} D line cores, the density of perturbers such as neutral hydrogen is low but not negligible. These collisions impact the scattering polarization signals both through the frequency redistribution discussed in the subsection above and by relaxing atomic level polarization (i.e., collisional depolarization).
In the absence of a magnetic field, the RT code used in step 2 of the calculations (see Sect.~\ref{sec::formulation}) can account for the depolarizing effect of the collisions that induce transitions between states that belong to the same HFS level, making use of the redistribution matrices presented in Appendix~C.7 of ABT22, in which the impact of depolarizing collisions is quantified by the $K$-multipole depolarizing rates $D^{(K)}(J_u, F_u)$. 

The intensity profiles are not impacted by the depolarizing effect of collisions and thus they are not shown here. Their impact on the D${}_2$ $Q/I$ profiles is illustrated in Fig.~\ref{fig::collisions}, 
which shows a comparison between the calculations setting $D^{(K)} = 0$ for all $K$ values and accounting for depolarizing collisions as follows. 
For all levels with $F \ge 1$, which can in principle carry alignment, the rates $D^{(2)}$ were estimated using the approximate relation $D^{(2)} = 0.5 \, \Gamma_E$. The rates $D^{(1)}$ for the same levels were obtained using a relation derived under the assumption that these collisions can be described through a Van der Waals interaction (see Sect.~7.13 of LL04). 
These depolarizing collisions have a noticeable impact on the D${}_2$ $Q/I$ profile, both in the dips in the near wing and in the peak in the line core, where a decrease of about $8\%$ is found. 
Although not shown in the figure, it can be verified that a comparable relative depolarization occurs in the line-core peaks of D${}_1$. 

It should be noted that the relations between the rates of depolarizing collisions and the elastic collisional rates considered in this work represent a rough estimate, and the calculation of the latter rates may be subject to further inaccuracies. In addition, the redistribution matrix formalism considered in this work can only account for the depolarizing effect of collisions in the absence of magnetic fields.\footnote{In the presence of magnetic fields that break the degeneracy of the $f$ states that belong to the same HFS level, the collisional coupling between such states precludes the analytical solution of the master equation \citep[for details, see][]{Bommier17}. 
As a result, the corresponding redistribution matrix cannot be derived.} Moreover, the depolarization due to collisionally induced transitions between states that belong to the same term, but which have different $J$ or $F$ numbers, cannot be taken into account even in the nonmagnetic case, even though they could be expected to have an impact comparable to that of the transitions within the same $F$ level \citep[see][]{Bommier17}. 
Therefore, despite the clear impact that depolarizing collisions have on the scattering polarization signals, they will not be taken into account in the rest of this work. As such, the scattering polarization amplitude of the K~{\sc{i}} D lines is somewhat overestimated. 
A more rigorous analysis of the impact of the depolarizing collisions in the K~{\sc{i}} D lines is left for a forthcoming publication. 

\subsection{The impact of the $J$- and $F$-state interference}
\label{sec::FSHFS}
\begin{figure*}[!t]
\centering
\includegraphics[width = 0.975\textwidth]{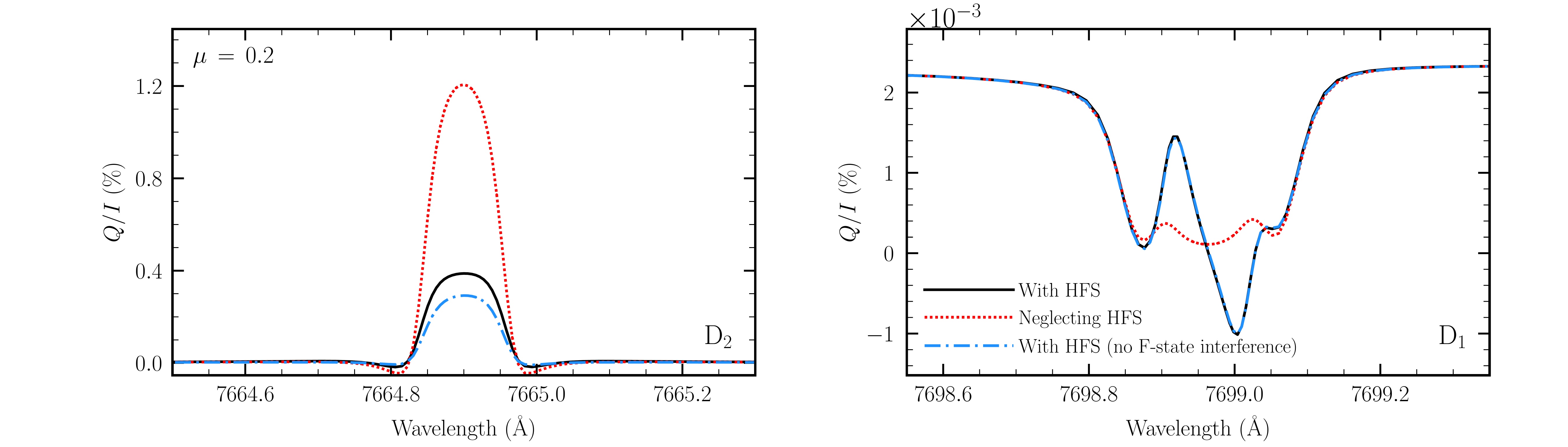}
\caption{Stokes $Q/I$ profiles as a function of wavelength in the spectral ranges centered on the K~{\sc i} D${}_2$ (left panel) and D$_{1}$ (right panel) lines. The RT calculations were carried out in the absence of a magnetic field. 
A comparison is shown between the calculations considering a two-term atomic model with HFS, both taking $F$-state interference into account (black solid curves) and neglecting it (blue dashed-dotted curves), and the calculation for a two-term atomic model without HFS (red dotted curves). } 
  \label{fig::hfs} 
\end{figure*}

This subsection is focused on the impact of $J$- state and $F$-state interference (see Sect.~\ref{sec::BasEq}) on the scattering polarization patterns of the K~{\sc{i}} D-lines. 
First, the impact of $J$-state interference is investigated through a comparison of the $Q/I$ patterns obtained through three different calculations, 
which in all cases accounted for HFS and $F$-state interference. The results of the comparison are highlighted in Fig.~\ref{fig::finestruc}. 
Two of the calculations were carried out considering two-term atomic models, accounting for $J$-state interference (black curves)
and neglecting it by artificially setting to zero the terms in the redistribution matrix that couple states that belong to different $J$ levels (blue curves; see Appendix~C.7 of ABT22 for details). 
Finally, calculations for the D${}_1$ and D${}_2$ lines were carried out separately by considering two-level models, for which $J$-state interference cannot be included by definition but $F$-state interference was taken into account (red curves).  
For both lines, a perfect agreement is found between the results of the two-level calculations and the two-term ones in which $J$-state interference was artificially neglected. Fully accounting for this interference has no appreciable impact on the D${}_2$ scattering polarization and only a small one on the wings of D${}_1$. Regardless, the rest of the calculations presented in this work considered a two-term atomic model in which $J$-state interference was taken into account, unless otherwise noted. 

The impact of HFS and $F$-state interference is highlighted in Fig.~\ref{fig::hfs}, which presents a comparison between the $Q/I$ profiles obtained by considering two-term atomic models with HFS, taking $F$- and $J$-state interference into account (black curves) and neglecting $F$-state interference (blue curves). The latter was implemented in a manner similar to what was discussed in Appendix~C.7 of ABT22, but only neglecting the interference between the $F$-levels that belong to the same $J$-level. The same figure also contains the results of the calculations considering a two-term atomic model in which the HFS is neglected by artificially setting the nuclear spin $I$ to zero, while still accounting for $J$-state interference (red curves). 
The intensity profiles, which are not impacted by the inclusion of HFS or by $F$-state interference, are not shown. 

\begin{figure*}[!t]
\centering
\includegraphics[width = 0.95\textwidth]{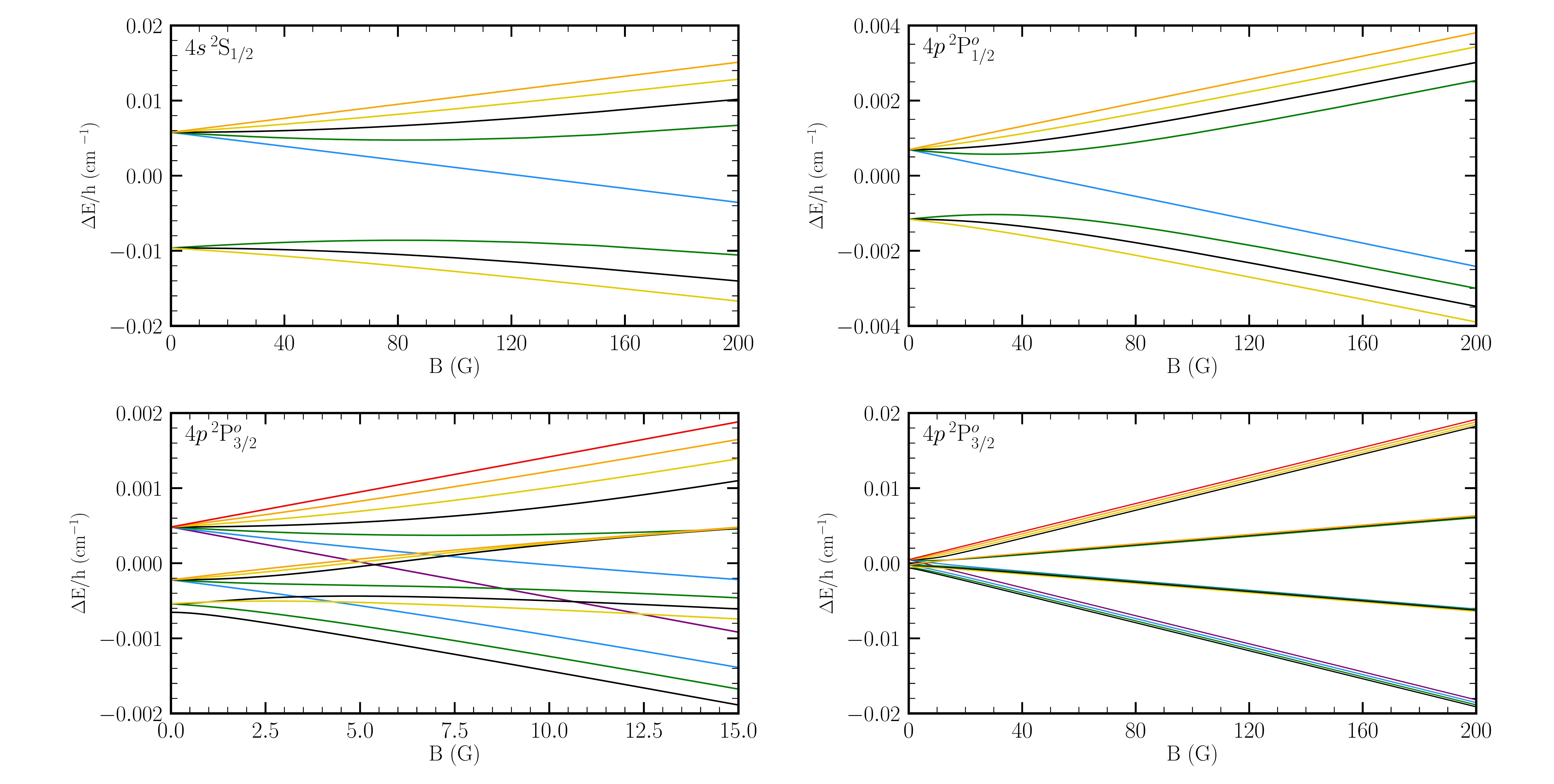}
\caption{Energies of the magnetic $f$ states associated with the $4s^2 \mathrm{S}_{1/2}$ (upper left {panel}), $4p^2 \mathrm{P}^{\mathrm{o}}_{1/2}$ (upper right panel), and $4p^2 \mathrm{P}^{\mathrm{o}}_{3/2}$ (the two lower panels) FS levels of K~{\sc{i}}, relative to their corresponding baricenters. A range of magnetic field strengths between $0$ and $200$~G is selected in all panels except the lower left, in which a smaller range is displayed to highlight various level crossings. The colors of the curves indicate the magnetic quantum number $f$ of each state, including $-3$ (purple), $-2$ (blue), $-1$ (green), $0$ (black), $1$ (yellow), $2$ (orange), and $3$ (red). }
	\label{fig::EnergyDiag} 
\end{figure*} 
The $Q/I$ profile of D${}_2$ is substantially depolarized by the inclusion of HFS. When not taking it into account, the scattering polarization amplitude increases by roughly a factor $3$. On the other hand, by including HFS but neglecting $F$-state interference, the resulting scattering polarization is further depolarized, and it presents a signal roughly $4$ times smaller than the one obtained without HFS. 
This behavior is consistent with the theoretical HFS depolarization factors of the scattering phase matrix $\bigl[D_{K}\bigr]_{\mbox{\scriptsize HFS}}$ for $K = 2$ (see Sect~10.22 of LL04), related to atomic alignment. 
When accounting for HFS but assuming that the hyperfine splitting between $F$-levels is zero, the depolarizing effect of HFS vanishes and $\bigl[D_{2}\bigr]_{\mbox{\scriptsize HFS}} = 1$. In this case, the various $F$ levels are degenerate, and therefore the quantum interference between them is maximum. When there is a splitting between the various $F$ levels, the interference between them decreases, causing a depolarization. This depolarization increases with the hyperfine splitting until the splitting is much larger than the natural width of the line, at which point the quantum interference between $F$ levels is effectively zero. 
At this limit, the depolarization for the K~{\sc i} D${}_2$ line is given by $\bigl[D_{2}^\infty \bigr]_{\mbox{\scriptsize HFS}} = 0.27$. 
When considering the real HFS splitting of the upper level of the line, the theoretical value of $\bigl[D_{2}\bigr]_{\mbox{\scriptsize HFS}}$ rises to $0.36$. The polarization profiles shown in Fig.~\ref{fig::hfs} are slightly more depolarized by HFS than would be predicted from these factors alone, as a consequence of transfer effects that reduce the anisotropy of the radiation field in the solar atmosphere, thereby reducing the $Q/I$ signal further.  

Despite the small scattering polarization amplitude of the D${}_1$ line, HFS has an obvious impact on its linear polarization pattern. Indeed, the antisymmetric profile around the line core is no longer found when neglecting HFS; in this case the depolarized feature typical of intrinsically unpolarizable absorption lines is produced instead. This is fully consistent with the physical origin of the antisymmetric signal around the line core discussed at the beginning of Sect.~\ref{sec::nonmag}. Although not shown here, it can also be verified that this signal is mainly a consequence of the hyperfine splitting of the lower term, which is roughly one order of magnitude larger than that of the upper term. 
Indeed, neglecting the $F$-state interference between states of the upper term has no appreciable impact on the D${}_1$ $Q/I$ pattern.  

\section{The magnetic sensitivity of the K~{\sc{i}} D lines}
\label{sec::magnetic}
Ultimately, the magnetic field modifies the intensity and polarization profiles by inducing energy shifts in the states that are involved in the transitions that contribute to the considered spectral line. 
On one hand, these energy shifts give rise to frequency shifts between the various components of the spectral line, leading to a displacement of the baricenters of the $\sigma$ and $\pi$ components of the line; this produces the intensity and polarization signatures commonly associated with the Zeeman effect. On the other, the magnetically induced changes in the energy of the atomic states also affect the atomic level polarization, which is responsible for scattering polarization. 

Moreover, in the presence of a magnetic field that is strong enough to produce a splitting within a given $F$ level comparable to the separation between the various $F$ levels, the eigenstates of the Hamiltonian for an atomic system with HFS no longer coincide with those of atomic (electronic plus nuclear) total angular momentum $\mathbf{F}$. Thus, the magnetic field introduces a mixing between states characterized by different $F$, which is thus no longer a good quantum number of the system. Due to this mixing, the dependency of the eigenstates of the Hamiltonian (for which the $f$ remains a good quantum number) on the magnetic field strength deviates from linearity. This phenomenon is referred to as the PB effect for HFS (or Back-Goudsmit) effect. 
In principle, the PB effect would also induce a mixing between states with different $J$ when the magnetic splitting within $J$ levels becomes comparable to the separation between the various $J$ levels. However, the energy separation between the $J$ (or FS) levels responsible for the K~{\sc{i}} D lines (see Appendix~\ref{sec::AppAtomicProps}) is large enough that, in the presence of magnetic fields with strengths typically encountered in the Sun, $J$ can be considered a good quantum number of the Hamiltonian. Thus, the various eigenstates of the Hamiltonian can be safely associated with a specific $J$ level.   

In Fig.~\ref{fig::EnergyDiag}, the energies of the various $f$ states of the ground level ($4s \,{}^2 \mathrm{S}_{1/2}$) and the upper levels of both the D${}_1$ ($4p \, {}^2 \mathrm{P}^{\mathrm{o}}_{1/2}$) and D${}_2$ ($4p \,{}^2 \mathrm{P}^{\mathrm{o}}_{3/2}$) lines are shown as a function of the field strength, relative to their corresponding baricenters. 
The ground level and the upper level of the D${}_1$ line have the same quantum numbers and their various $f$ states have a qualitatively similar magnetic dependency. 
However, the hyperfine splitting is smaller for the upper level and, as a result, deviations from linearity, especially in the form of repulsions between $f$ states,\footnote{The crossing of the energies of states with different $f$ numbers and the apparent repulsion between states with the same $f$ may produce signatures in the scattering polarization that are often referred to as level-crossing and antilevel-crossing effects, respectively \citep[e.g.,][]{Bommier80}.}  
are appreciable for weaker magnetic fields, as low as $15$~G. For magnetic fields stronger than $100$~G, a linear dependency with the field strength is regained for the upper level of D${}_1$ as the transition from the incomplete to the complete PB effect regime is made.  
For the ground level, deviations from linearity are not clearly appreciable until field strengths of $\sim\!40$~G are considered and the complete PB effect regime is not reached until well beyond $200$~G. 
On the other hand, the hyperfine splitting for the upper level of D${}_2$ is even smaller than for the other FS levels. 
Level crossings can be identified in $4p^2 \mathrm{P}^{\mathrm{o}}_{3/2}$ for field strengths as low as $3$~G, and repulsions can be appreciated in the presence of even weaker fields. The complete PB effect regime is reached at around $40$~G. 

\begin{figure*}[!t]
\centering
\includegraphics[width = 0.975\textwidth]{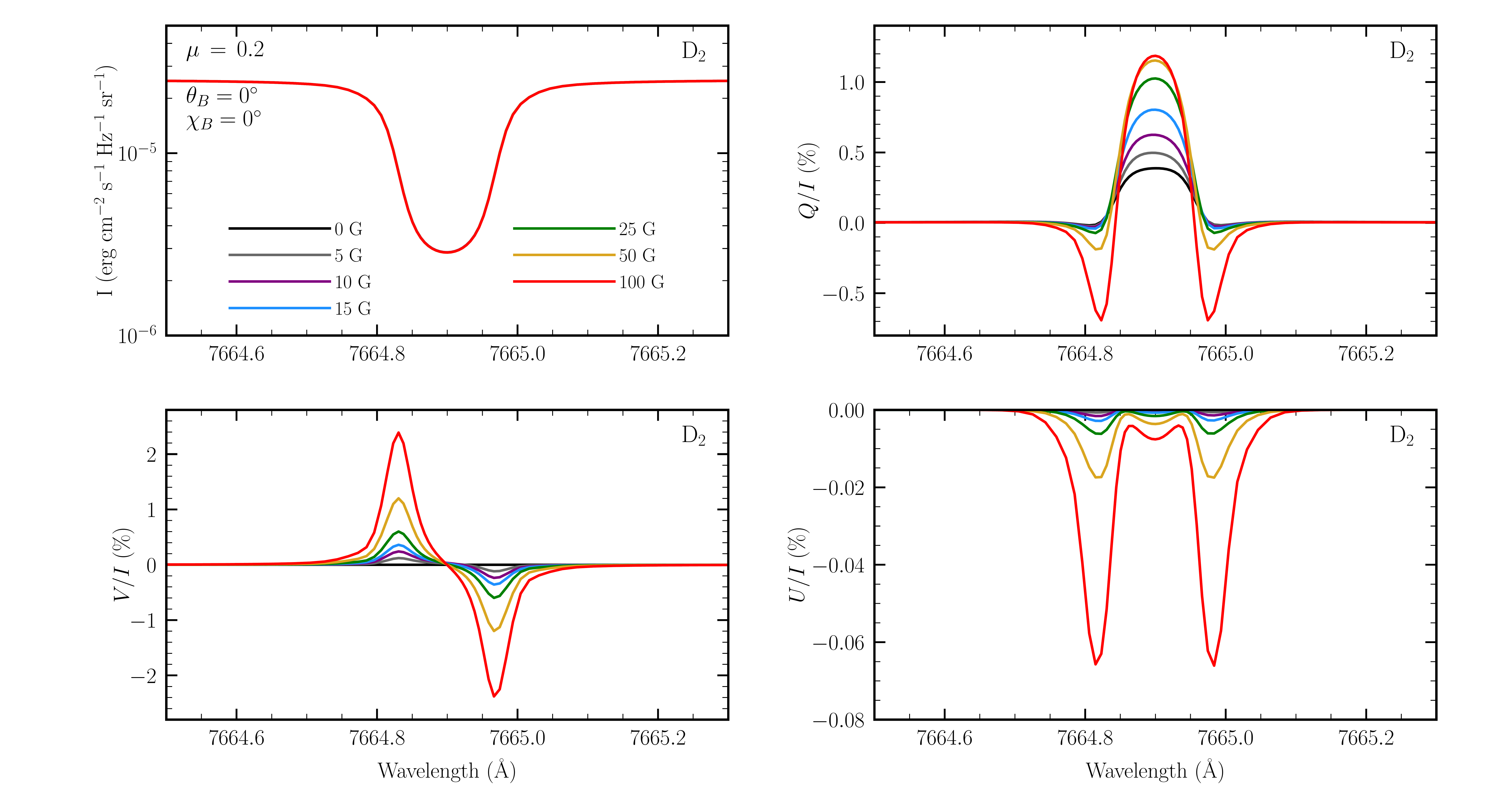}
	\caption{Stokes $I$ (upper left panel), $V/I$ (lower left panel), $Q/I$ (upper right panel), and $U/I$ profile (lower right panel) as a function of wavelength, calculated in the presence of a vertical ($\theta_B = 0^\circ$) magnetic field. The various colored curves represent different field strengths (see legend). The figures in this subsection only show the spectral range centered on the K~{\sc i} D${}_2$ line. For this figure, and the rest of those presented in this work, a two-term atomic model with HFS is considered.} 
	\label{fig::StokesVerticalSample} 
\end{figure*}

\subsection{The influence of the magnetic field on the D${}_2$ scattering polarization}
\label{sec::MagScatPol} 
The influence of the magnetic field on the scattering polarization patterns of the K~{\sc{i}} D${}_2$ line is analyzed in this subsection. 
\begin{figure*}[!t]
 \centering
\includegraphics[width = 0.975\textwidth]{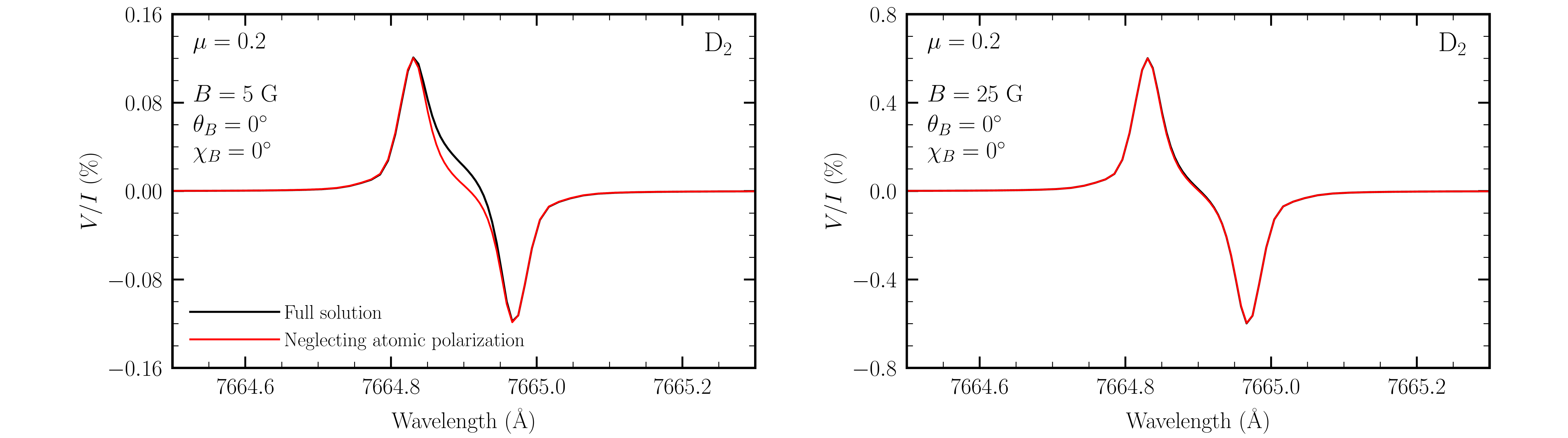}
 	\caption{Stokes $V/I$ profiles as a function of wavelength, shown in a range centered on the D${}_2$ and calculated in the presence of vertical ($\theta_B = 0^\circ$) magnetic fields of $5$ (left panel) and $25$~G (right panels). The black curves represent the results of RT calculations considering a two-term atomic model with HFS. The red curves represent calculations in which, additionally, all components of the radiation field tensor other than $J^0_0$ were set to zero, which effectively suppresses atomic level polarization fully (see text). }  
 	\label{fig::V_AlignToOrien} 
 \end{figure*}
\begin{figure*}[!t]
\centering
\includegraphics[width = 0.975\textwidth]{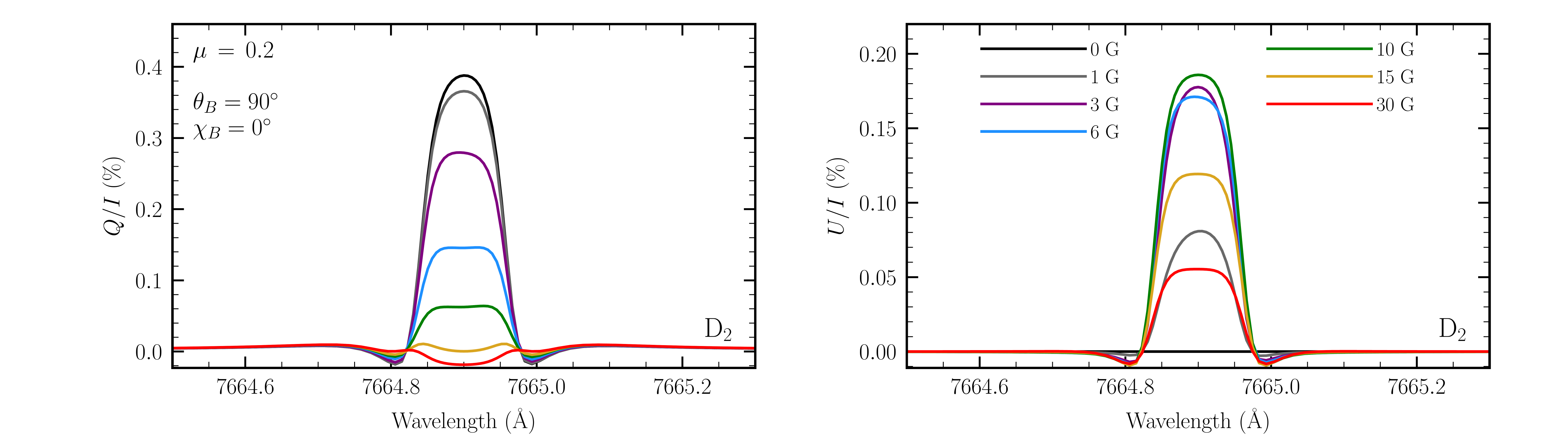}
	\caption{Stokes $Q/I$ (left panel), and $U/I$ (right panel) profiles as a function of wavelength, calculated in the presence of horizontal ($\theta_B = 90^\circ$) magnetic fields, contained in the plane defined by the local vertical and the LOS ($\chi_B = 0^\circ$). The various colored curves represent different field strengths (see legend). }
	\label{fig::StokesHorizontalSample} 
\end{figure*}
The discussion is focused on the synthetic Stokes profiles of this line, obtained for three different geometries for the magnetic field. In all cases, the fields are taken with the same strength and orientation at all height points of the model. 

\subsubsection{{Vertical magnetic fields}}
\label{sec::ResBvert}
The D${}_2$ line profiles that were obtained in the presence of vertical magnetic fields ($\theta_B = 0^\circ$) with various strengths between $0$ and $100$~G are shown in Fig.~\ref{fig::StokesVerticalSample}. 
The reader is referred to Figs.~\ref{fig::D2v00s} and \ref{fig::D1v00s} for the profiles obtained in the presence of stronger magnetic fields with the same geometry, for the D${}_2$ and D${}_1$ lines, respectively. The response function of the D${}_2$ line for a vertical magnetic field is presented in Fig.~\ref{fig::RespVert} in Sect.~\ref{sec::AppResponse}.  
The intensity profiles are insensitive to magnetic fields in the range of strengths considered here and therefore are not shown in the rest of figures in this section. 
 
The $Q/I$ amplitude is found to increase with field strength, but this trend begins to halt beyond $25$~G and only a small increase is found between $50$ and $100$~G. 
The magnetic enhancement of the scattering polarization signal in lines produced by atomic systems with substantial hyperfine splitting was  
already identified by \cite{TrujilloBueno+02}. 
If the magnetic field is vertical, and the incident radiation field is axially symmetric around the same direction, it only modifies the scattering polarization through the interference between states with the same quantum number $f$ (as can be seen from the expressions in Appendix~C.4 of ABT22). 
Thus, the magnetic enhancements are a consequence of the aforementioned repulsion between such states and of the ensuing change in the interference between them (this is an example of an antilevel-crossing effect). 

Because vertical magnetic fields do not break the axial symmetry of the problem, the scattering of anisotropic radiation does not give rise to $U/I$ signals in this case. The peaks in the wings of both the $Q/I$ and $U/I$ profiles are instead signatures associated with the Zeeman effect and are sensitive to the transverse component of the magnetic field. For stronger magnetic fields (see Fig.~\ref{fig::D2v00s}), 
the line-core $Q/I$ and $U/I$ are also dominated by such Zeeman-type signals. 

The antisymmetric circular polarization profiles, whose amplitude increases with the longitudinal component of the magnetic field, are also signatures associated with the Zeeman effect. 
Although it is not shown in this section, it was verified that the thermal contribution to the line emissivity (see Eq.~\eqref{eq::EmisLineTherm}) has no appreciable impact on the scattering polarization signals, but it does contribute to the Zeeman-type linear and circular polarization signals. In the presence of relatively weak magnetic fields, it is particularly important to take it into account when modeling $V/I$.  

For the geometry under consideration, the circular polarization of D${}_2$ may be impacted to a certain degree by the alignment-to-orientation (AtO) conversion mechanism \citep[see][also LL04]{Kemp+84}. This is apparent from the asymmetries found in the $V/I$ profile close to the line core in the presence of magnetic fields of around $5$~G, shown in the left panel of Fig.~{\ref{fig::V_AlignToOrien}}. However, even in this case the peak and area asymmetries of its Stokes $V$ signals are no larger than $2\%$ and $3\%$, respectively. 
Additional calculations were carried out in which all the components of the radiation field tensor (e.g., Sect.~5.11 of LL04) other than $J^0_0$ were artificially set to zero at each iteration in step 2 of the RT problem (see Sect.~\ref{sec::NumScheme}), which effectively sets the atomic level polarization to zero. 
As can be seen from the figure, the abovementioned asymmetry is no longer found in the core region in this case, which strongly indicates that it originates from the AtO mechanism. On the other hand, when considering stronger magnetic fields such that the complete PB effect regime is reached in the upper level of D${}_2$, the circular polarization pattern is fully dominated by Zeeman-type signatures, and asymmetries are not appreciable (see right panel). 
Asymmetries in the circular polarization of the Na~{\sc{i}} D${}_2$ line were also reported in the theoretical investigation of \cite{Sampoorna+19}, in which the line was modeled as a two-level atom with HFS. They were also identified as signatures of the incomplete PB effect and were found to be considerable in the presence of fields with strengths of up to roughly $30$~G.  

\subsubsection{{Horizontal magnetic fields}} 
\label{sec::ResBhor}
Horizontal magnetic fields contained in the plane defined by the local vertical and the LOS (i.e., $\theta_B = 90^\circ$ and $\chi_B = 0^\circ$), with strengths up to $30$~G, are considered here. The resulting linear polarization profiles are shown in Fig.~\ref{fig::StokesHorizontalSample}. 
The response function of the D${}_2$ line for a magnetic field with this geometry is presented in Fig.~\ref{fig::RespHoriz} in Sect.~\ref{sec::AppResponse}. 
For the considered field strengths, the transverse component is rather small and the linear polarization signals associated with the Zeeman effect are not appreciable. However, in the presence of stronger magnetic fields, such signatures become apparent in the linear polarization profile of both D lines (see Figs.~\ref{fig::D2v900s} and \ref{fig::D1v900s}).  

Circular polarization signals associated with the Zeeman effect are also found for this geometry. However, the $V/I$ profiles are not shown in the figure; a discussion on such signals, considering geometries for which the longitudinal component of the magnetic field is large, can be found in the dedicated Sect.~\ref{sec::circpol}. 
No appreciable asymmetries are found around the core region of the $V/I$ profile, even for field strengths for which the upper $J$ level of D${}_2$ is well within the incomplete PB effect regime. In principle, one may have expected the AtO mechanism to have a greater impact than in the presence of the vertical magnetic fields discussed above (e.g., Sect.~10.20 of LL04). The lack of apparent asymmetries for this geometry can be attributed to the fact that the horizontal magnetic field strongly reduces the overall atomic alignment, thereby reducing the impact of this conversion mechanism. Its impact is further masked by the fact that the circular polarization signals associated with the Zeeman effect, which are proportional to the longitudinal component of the magnetic field, have a considerably larger amplitude.  
\begin{figure*}[!t]
\centering
\includegraphics[width = 0.975\textwidth]{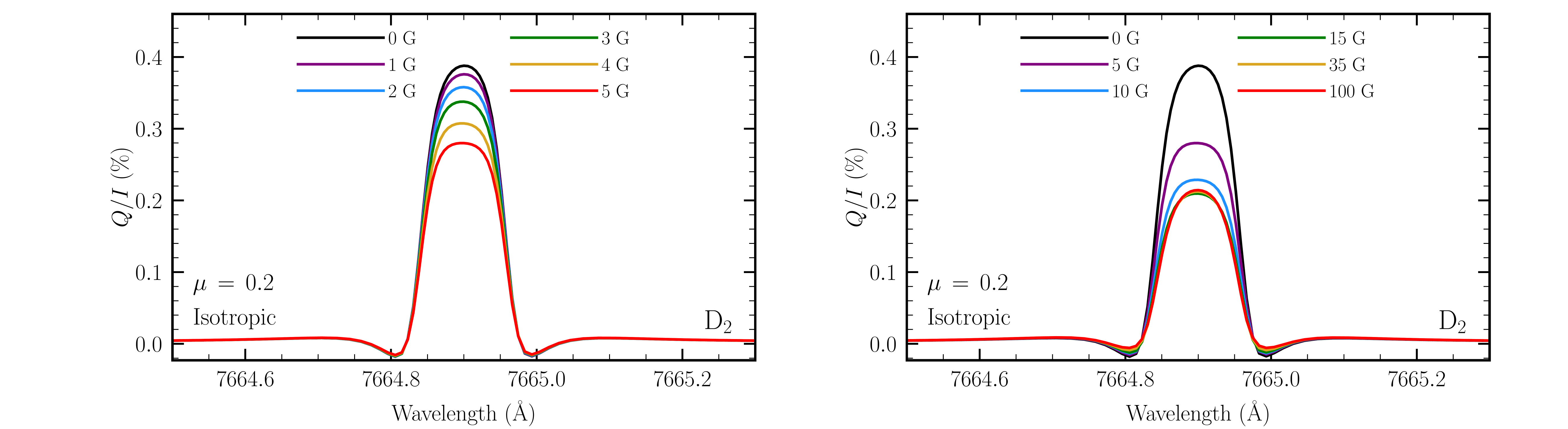}
\caption{Stokes $Q/I$ profiles as a function of wavelength, calculated in the presence of microstructured 
 and isotropic magnetic fields, with strengths up to $5$~G (left panel) and $100$~G (right panel). 
The various colored curves represent the different field strengths (see legend).}
 \label{fig::StokesIsotropicSample}
\end{figure*}

For the relatively weak magnetic fields considered in Fig.~\ref{fig::StokesHorizontalSample}, the $Q/I$ signal in the D${}_2$ core decreases progressively as the magnetic field increases, in contrast to the enhancement found in the presence of a vertical magnetic field.
Because the magnetic field is inclined with respect to the symmetry axis of the radiation field, the scattering polarization is also affected by the interference between states of the upper term such that $|\Delta f| = |f_u - f_u^\prime| \ne 0$.  
The impact of the magnetic field on such interference produces a decrease in the scattering polarization amplitude. 
 
The asymmetry introduced in the problem by the magnetic field gives rise to a $U/I$ signal. 
Its amplitude increases for field strengths up to $3$~G, then decreases until about $6$~G, and then increases again until reaching roughly $10$~G. This nonmonotonic behavior is a characteristic of level-crossing effects (see also  Appendix~\ref{sec::AppHanleDiag}). 
Indeed, the field strengths at which the second increase in the $U/I$ signal is found {correspond to the crossing of two states having} $f = -2$ and $f = 0$. 
As the magnetic field further increases, the energy levels of the various states separate further and their interference decreases, and thus the total scattering polarization amplitude continues to decrease, both in Stokes $Q/I$ and $U/I$, eventually reaching a saturation value when the complete PB effect regime is entered. 

\subsubsection{{Microstructured and isotropic magnetic fields}}
\label{sec::ResBiso}
In addition to deterministic magnetic fields (i.e., those with a single orientation at any given point on the spatial grid), the numerical code used for the step-2 calculations (see Sect.~\ref{sec::NumScheme}) can consider microstructured magnetic fields (i.e, those whose orientation changes over scales smaller than the mean free path of the line's photons), as discussed in Appendix~C.6 of ABT22. 
The Stokes $Q/I$ profiles obtained from calculations considering microstructured fields with an isotropic distribution of orientations and strengths up to $100$~G are shown in Fig.~\ref{fig::StokesIsotropicSample}. 
In this scenario, the magnetic field has no net longitudinal component and thus the $V/I$ signal remains zero. Because the magnetic field does not introduce any axial asymmetry, no $U/I$ signal is produced either and thus only $Q/I$ is shown in the figure. 
For field strengths up to roughly $15$~G, the $Q/I$ signal of the D${}_2$ line decreases as the magnetic field becomes stronger. The $Q/I$ then reaches an amplitude minimum, which is slightly above half the nonmagnetic value. 
As the field strength increases further, the $Q/I$ increases by a small amount, reaching a plateau as the upper level of the D${}_2$ enters the complete PB effect regime. The scattering polarization amplitude remains almost constant beyond $50$~G, which is a similar phenomenon to the well-know Hanle saturation that has been extensively studied in two-level atoms (e.g., Sect.~5.12 of LL04).  

The same behavior was theoretically described for the Na~{\sc{i}} D${}_2$ line in Sect.~10.22 of LL04, in which the quantity $\bigl[M_K(B\to\infty)\bigr]_{\mbox{\scriptsize{hfs}}}/\bigl[M_K(0)\bigr]_{\mbox{\scriptsize{hfs}}}$ was introduced. This ratio characterizes the magnetically induced depolarization of the components of the emission vector with a specific value of the quantum number $K$ in the complete PB effect regime, relative to its nonmagnetic value. 
For $K = 2$, this ratio is a good indication of the decrease in the linear polarization signal for the K~{\sc{i}} D${}_2$ line. Indeed, for the quantum numbers, FS and HFS splitting, and radiative line broadening of the K~{\sc{i}} D${}_2$ line, this ratio is roughly $0.55$, which is consistent with the results shown in Fig.~\ref{fig::StokesIsotropicSample}. 

Interestingly, the $Q/I$ amplitude obtained at field strengths beyond $50$~G coincides with the one found when neglecting HFS in the presence of  isotropic magnetic fields with the same strengths (not shown). As discussed above, the $Q/I$ amplitude of the D${}_2$ line obtained in the absence of a magnetic field is substantially larger when HFS is neglected (see Fig.~\ref{fig::hfs}). Thus, in the case without HFS, a microstructured and isotropic magnetic field at saturation reduces the amplitude by the well-known factor $0.2$ \citep[e.g.,][]{TrujilloBuenoMansoSainz99} instead of by $0.55$. Indeed, the product of the (nonmagnetic) depolarization factor due to HFS discussed in Sect.~\ref{sec::FSHFS} and the magnetic depolarization factor accounting for HFS discussed in the previous paragraph is $0.36 \times 0.55 \simeq 0.2$. 

\begin{figure*}[!t]
\includegraphics[width = 0.975\textwidth]{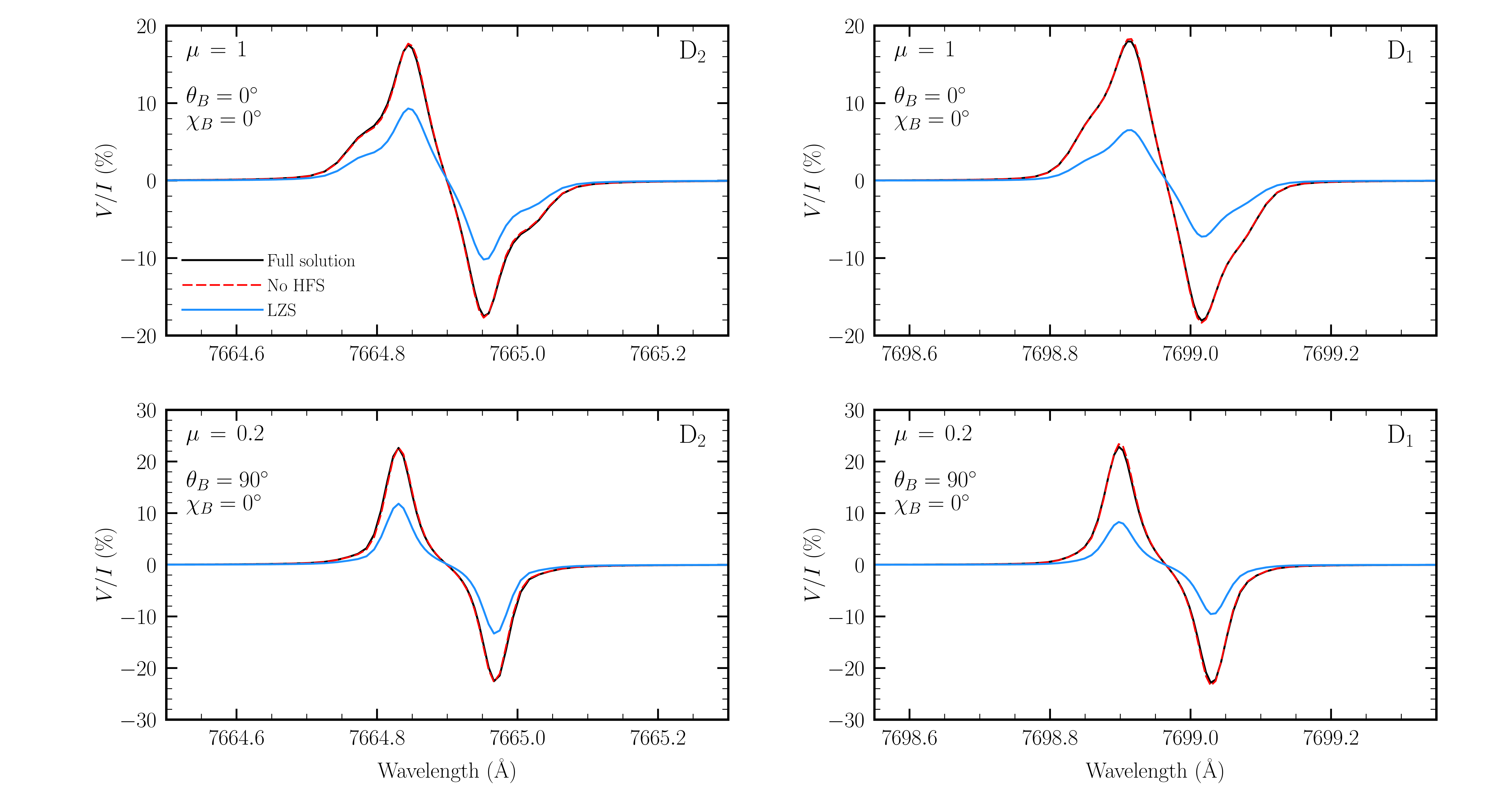}
\caption{Stokes $V/I$ profiles as a function of wavelength, for the spectral ranges centered on the K~{\sc{i}} D${}_2$ (left panels) and D${}_1$ (right panels) lines, respectively. Upper panels: an LOS of $\mu = 1$ is considered in the presence of vertical magnetic fields. 
Lower panels: an LOS of $\mu = 0.2$ is considered in the presence of horizontal fields with azimuth $\chi_B = 0^\circ$. 
A field strength of $200$~G is considered for both geometries. The results of RT calculations accounting for HFS and accounting for the Paschen-Back (PB) effect 
(black solid curves) are compared to calculations neglecting HFS in the atomic model (red dashed-dotted curves) and calculations accounting for HFS but making the linear Zeeman splitting (LZS) approximation as described in the text (blue solid curves). } 
\label{fig::V_nHFS_LinZeem}
\end{figure*}

\subsection{Circular polarization signals due to the Zeeman effect}
\label{sec::circpol} 
This subsection is dedicated to analyzing the circular polarization signals the K~{\sc{i}} D${}_2$ and D${}_1$ lines. The amplitude of these signals scales linearly with the longitudinal component of the magnetic field up until strengths of a few hundred gauss; beyond this point, the magnetic splitting becomes comparable to the Doppler width and this linearity is lost (see Sect.~9.6 of LL04). Figures showing the profiles obtained in the presence of stronger fields can be found in Appendix.~\ref{sec::AppExtraFig}.   
The two geometries considered here were selected so that the magnetic field has a large longitudinal component. 
For the first geometry, an LOS with $\mu = 1$ and a vertical magnetic field was considered, whereas for the other an LOS with $\mu = 0.2$ and a horizontal magnetic field with $\chi_B = 0^\circ$ was taken. 
In both cases, a field strength of $200$~G was selected. 
The resulting $V/I$ profiles for the D${}_1$ and D${}_2$ lines are indicated with the black curves in Fig.~\ref{fig::V_nHFS_LinZeem}.  

When considering a vertical magnetic field and an LOS with $\mu = 1$, secondary $V/I$ peaks are found farther into the wings than the main peaks, with the same sign. By carrying out test RT calculations under the same conditions, but setting to zero all components of the radiation field tensor except for $J^0_0$ at each iteration as discussed in Sect.~\ref{sec::ResBvert} (not shown for this geometry), it was verified that these secondary lobes are unaffected by the other components \citep[in contrast to what was found for the Mg~{\sc{ii}} $k$ line in][where the $J^1_Q$ radiation field tensors were found to have an appreciable impact in the wings]{AlsinaBallester+16}. 
This implies that these lobes are not produced by the AtO mechanism. On the other hand, test calculations carried out without accounting for the impact of the magnetic field on the thermal line contribution to the emissivity of Eq.~\eqref{eq::EmisLineTherm} greatly overestimated the amplitude of these secondary peaks. 
In addition, it was verified that PRD effects only have a minimal impact on the circular polarization profile through a comparison with CRD calculations, although these results are not explicitly shown in this paper either. 

The computational cost of evaluating the emission vector can be reduced substantially by neglecting HFS (compare the expressions for the redistribution matrices with and without HFS, which are given in Appendices~C.4 and~C.8 of ABT22, respectively). Although this approach is found to be unsuitable for modeling the scattering polarization signals of the D lines, it allows for an accurate modeling of their intensity profiles (see Sect.~\ref{sec::FSHFS}). Here, its suitability for modeling the $V/I$ signals is evaluated by comparing the results of RT calculations in which HFS is taken into account and neglected. Such a comparison is shown in Fig.~\ref{fig::V_nHFS_LinZeem} for the two abovementioned geometries and taking a field strength of $200$~G, for which the complete PB effect regime for HFS is reached in the upper FS levels of both lines. 
The impact of HFS is barely appreciable in the $V/I$ signals and thus, for most practical purposes, it can be neglected when modeling the circular polarization signals of these lines. 
On the other hand, it is worth recalling that HFS is required to reproduce the slight asymmetries in the D${}_2$ $V/I$ profile due to the AtO mechanism discussed in Sect.~\ref{sec::ResBvert}. Calculations identical to the ones discussed here, but considering a field strength of $50$~G, were also carried out. In this case, the impact of HFS on the $V/I$ profiles was likewise found to be negligible. 

Alternatively, when accounting for HFS, one may be tempted to simplify the numerical calculations by making the so-called linear Zeeman splitting (LZS) approximation, in which the energy of all the magnetic sublevels is assumed to vary linearly with the field strength. 
This can be achieved in the RT calculations by artificially setting to zero the off-diagonal elements of the magnetic Hamiltonian that couple states with different $F$ or $J$ (see Appendix~C.1 of ABT22), thereby neglecting the mixing between states with different $F$ (or $J$) quantum numbers that characterizes the PB effect. 
A comparison of the $V/I$ profiles calculated considering the LZS approximation (see blue curves in Fig.~\ref{fig::V_nHFS_LinZeem}) and fully accounting for the PB effect (black curves) reveals the approximation to be unsuitable. Indeed, the LZS approximation yields a maximum amplitude that is too small by roughly a factor 2. 
Qualitatively similar results were found when considering a field with a strength of $50$~G instead of $200$~G (not shown in this paper). 
 
 \begin{figure*}[!t]
\centering
\includegraphics[width = 0.975\textwidth]{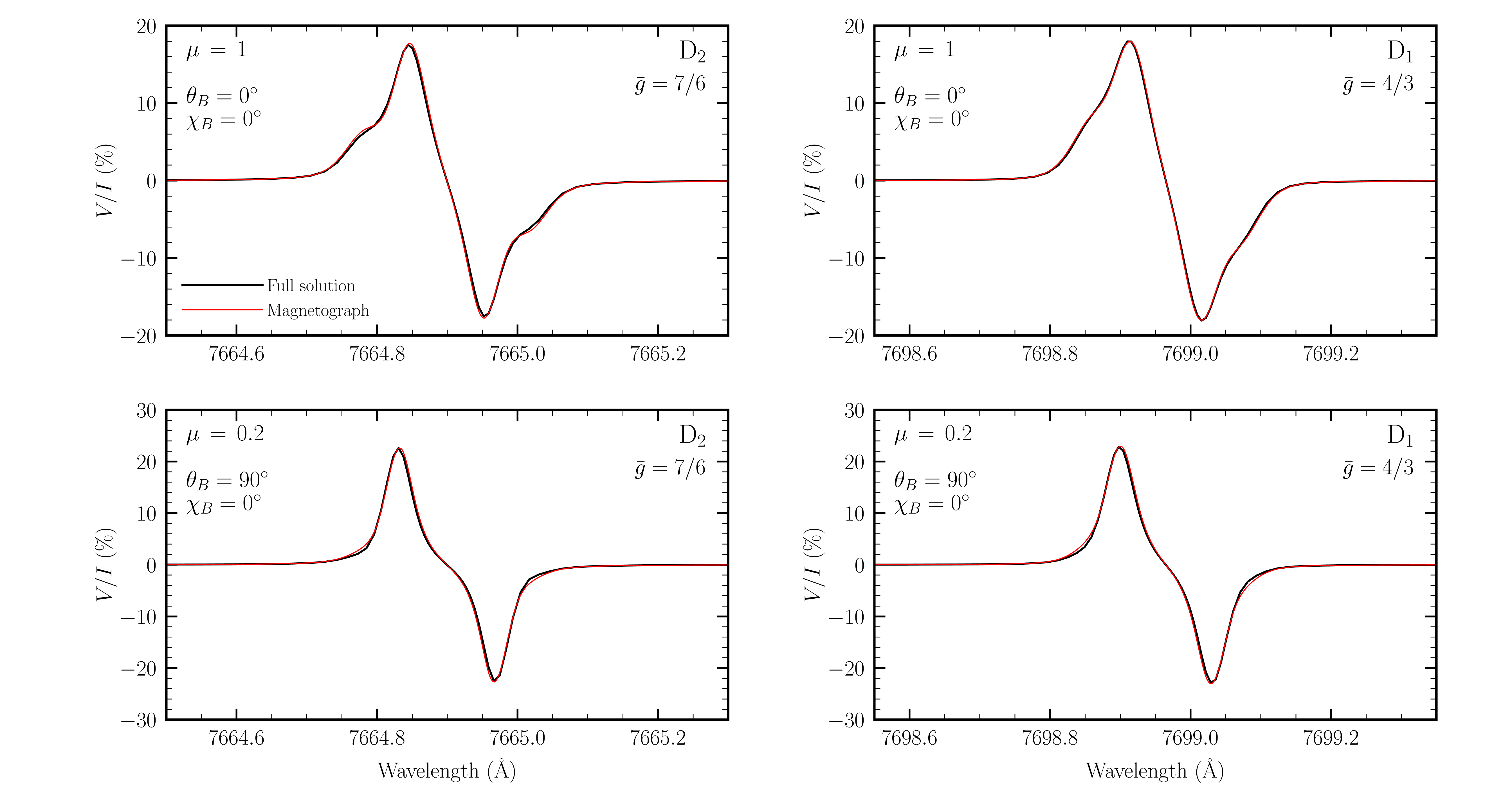}
\caption{Stokes $V/I$ profiles as a function of wavelength, for the spectral ranges centered on the K~{\sc{i}} D${}_2$ (left panels) and D${}_1$ (right panels) lines, respectively. 
Upper panels: an LOS of $\mu = 1$ is considered in the presence of a $200$~G vertical magnetic field. 
Lower panels: an LOS of $\mu = 0.2$ is considered in the presence of a $200$~G horizontal field with azimuth $\chi_B = 0^\circ$. 
The results of RT calculations considering an atomic model with HFS (black solid curves) are compared to those obtained through the magnetograph formula, in which the effective Land\'{e} factors $\bar{g}$ were obtained without accounting for HFS (red solid curves). The $\bar{g}$ values for the D${}_1$ and D${}_2$ lines are shown in the corresponding panels.} 
\label{fig::StokV_WFA_nHFS}
\end{figure*}
A common approach to obtain information on solar magnetic fields from observations of the circular polarization signals in spectral lines is the magnetograph formula (see Sect.~9.6 of LL04), which is sometimes referred to as the weak field approximation. 
This formula relates the circular polarization signal to the wavelength derivative of the intensity, with a factor that depends on the longitudinal component of the magnetic field and the effective Land\'e factor $\bar{g}$. 
Given that the centers of the D${}_1$ and D${}_2$ lines are separated by $\sim\!\!30$~\AA , this approach can be applied to each of them individually, effectively treating them as if they arise from distinct two-level atomic systems (see Sect.~\ref{sec::FSHFS}). 
Furthermore, bearing in mind that HFS has no significant impact on the $V/I$ signal, the effective Land\'e factors of the two lines (e.g., Sect.~3.3 of LL04) can be calculated without taking it into account. Under the assumption of L-S (or Russel-Saunders) coupling, the effective Land\'{e} factors for the D${}_1$ and D${}_2$ lines are $4/3$ and $7/6$, respectively. 
The profiles obtained through the application of the magnetograph formula as detailed here reproduce the results of the full RT calculation accounting for HFS remarkably well. This is illustrated in Fig.~\ref{fig::StokV_WFA_nHFS} for the two abovementioned geometries. 
Thus, valuable information on the longitudinal component of the magnetic field can be accessed through the application of the magnetograph formula. 
Finally, it must be recalled that its applicability is limited to the range of field strengths for which the magnetic splitting between $f$ states is much smaller than the width of the line profile (i.e., weak fields). Indeed, as the field strength increases beyond $\sim\!350$~G, the results obtained under the magnetograph formula deviate increasingly from those obtained with full RT calculations (see the figures in Appendix.~\ref{sec::AppExtraFig}). 

\begin{figure*}[!t]
\centering 
\includegraphics[width = 0.975\textwidth]{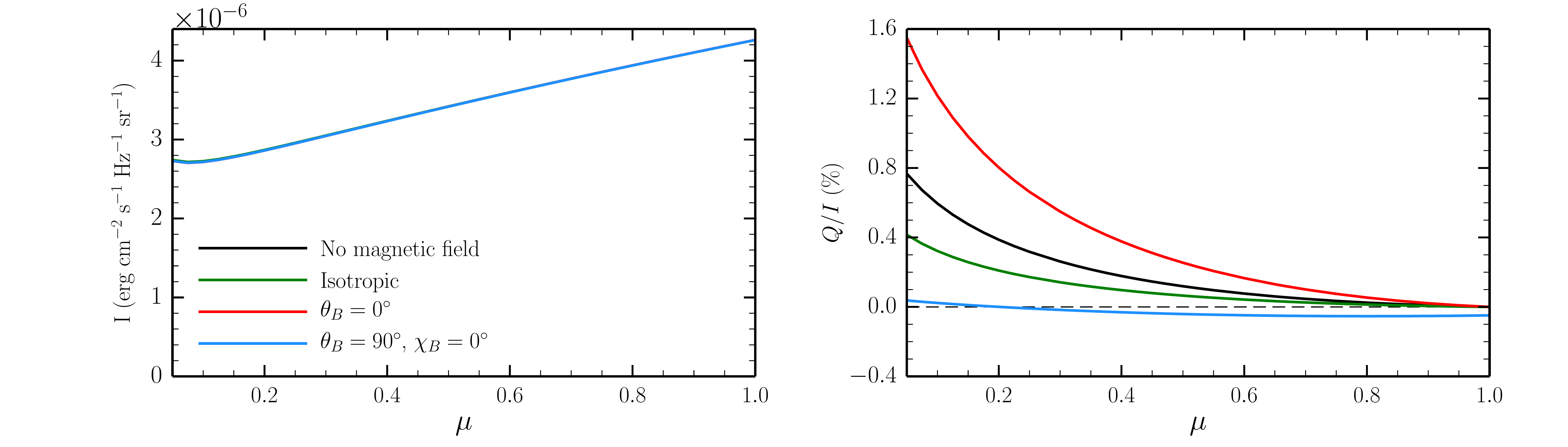}
\caption{Center-to-limb variation (CLV) of the Stokes $I$ (left panel) and $Q/I$ (right panel) signal at the wavelength $\lambda_0$ at the D${}_2$ line center. The colored curves correspond to the results of calculations in which $15$~G magnetic fields were considered, including a microturbulent and isotropic field (green curve) and deterministic magnetic fields, both vertical ($\theta_B = 0^\circ$; red curve) and horizontal and contained in the plane defined by the local vertical and the LOS ($\theta_B = 90^\circ$, $\chi_B = 0^\circ$; blue curve). The CLV obtained in the absence of a magnetic field (black curve) is also shown. The thin black dashed line indicates $Q/I = 0$.} 
\label{fig::CLV}
\end{figure*} 

\section{Center-to-limb variation at the D${}_2$ line center} 
\label{sec::CLV} 
In Sect.~\ref{sec::MagScatPol}, special attention was paid to the intensity and scattering polarization signals found at an LOS with $\mu = 0.2$. However, it may be of interest to consider also the radiation emerging closer to the disk center, especially for comparisons with observations. 
Here, the CLV of the intensity and $Q/I$ is investigated in the presence of magnetic fields of different orientations, assuming them to be constant with height. The results of calculations considering the FAL-C atmospheric model are shown in Fig.~\ref{fig::CLV} for a series of LOSs between $\mu=0.05$ and $1$, in steps of $0.05$, at an air wavelength that corresponds to the center of the D${}_2$ line ($\lambda_0 = 7664.89$~\AA ). 
The results of the nonmagnetic case are compared to those of the three magnetic field realizations investigated in Sect.~\ref{sec::MagScatPol}, taking a field strength of $15$~G in all cases.  
These realizations include two different deterministic magnetic fields, one horizontal within the plane defined by the local vertical and the LOS ($\theta_B = 90^\circ$ and $\chi_B = 0^\circ$) and one vertical ($\theta_B = 0^\circ$), as well as a microstructured and isotropic field. 

The intensity exhibits the same CLV in the four cases, which is consistent with the relative lack of magnetic sensitivity discussed in Sect.~\ref{sec::MagScatPol}. A clear limb darkening is found for D${}_2$ and a similar behavior is found at the center of the D${}_1$ line (not shown). This occurs because, for wavelengths close to the center of these lines, the intensity of the radiation field sharply decreases from the upper photosphere to the lower chromosphere, and the radiation emerging closer to the limb is generally sensitive to conditions at higher atmospheric regions.  

Closely related to the limb darkening is the fact that the degree of anisotropy of the radiation field increases with height at the same spatial regions. 
Both the height dependence of the radiation anisotropy and the increase of $I$ toward disk center contribute to the fact that $Q/I$ decreases with $\mu$ much faster than would be expected from the well-known $\sim\!\!(1 - \mu^2)$ trend \citep[e.g.,][]{TrujilloBueno03}, which is often assumed when considering slab models. 
It is also noteworthy that, although the $Q/I$ amplitudes strongly vary among the considered magnetic field realizations (as discussed in Sect.~\ref{sec::MagScatPol}), the relative variation of the amplitude with $\mu$ is quite similar for the four cases, at least under the assumption that the strength and orientation of the magnetic field is constant with height. 
Despite this, in the presence of a horizontal magnetic field, the linear polarization signal does not fall to zero at the disk center ($\mu = 1$) but instead becomes negative due to the Hanle effect for forward scattering \citep[e.g.,][also Sect.~5.13 of LL04]{TrujilloBueno01}. 
In any case, the fact that the relative CLV of the $Q/I$ signal is only weakly sensitive to the orientation of the magnetic field is an interesting finding from the diagnostic perspective. 
Used in combination with the information on the longitudinal component of the magnetic field, which is available through the circular polarization signals, the scattering polarization can provide constraints on the magnetic field vector and on the anisotropy of the radiation field. 

\section{Response functions for the magnetic field with scattering polarization and HFS}
\label{sec::AppResponse} 
The intensity and polarization signals of spectral lines often present their strongest sensitivity to different physical properties of the atmosphere at different depths. Indeed, \cite{QuinteroNoda+17} reported that the line-core intensity of the K~{\sc{i}} D lines is mainly sensitive to the temperature in the lower photosphere, whereas the sensitivities of the intensity to the LOS velocity and of $V/I$ to the magnetic field were found to be strongest in the upper photosphere. That investigation was focused on the so-called response functions, which quantify the sensitivity of a given  
{observable} (in that case the Stokes parameters) to a specific atmospheric property $X(\mathbf{r})$. 
In a plane-parallel one-dimensional atmospheric model such that its properties depend only on the height $z$, it is given by 
\begin{equation}
 I_{i}(\lambda) = \int_{-\infty}^{z_0} \! R_{i}^{X}(\lambda,z) \, X(z) \, \mathrm{d}z \, ,   
 \label{eqapp::StokesRespons}
\end{equation} 
where $I_i$ is the $i$-th Stokes parameter of the emergent radiation (see Sect.~\ref{sec::BasEq}), $z_0$ is the upper boundary of the atmospheric model,  and $R_i^X$ is the response function of $I_i$ to $X$. 
$I_i$ and $R_i^X$ also depend on the LOS, but this dependency is omitted for simplicity of notation. 
This section is focused on the response functions for the magnetic field in the spectral region around the D${}_2$ line, in which the impact of scattering polarization and HFS is taken into account for the first time. The response of $I_i$ to the magnetic field at height $z$ was determined by adding a perturbation to the magnetic field strength from the lower boundary of the problem up to that same height, such that  
\begin{equation}
 B(z^\prime) = \Delta B \, H(z - z^\prime) + B^{(0)} \, , 
 \label{eqapp::MagnPertur}
\end{equation}
where $H$ is the step function, which is zero for $z^\prime \!>\! z$ and unity otherwise. $\Delta B$ is the magnetic perturbation, and $B^{(0)}$ is the strength of the unperturbed magnetic field. Following \cite{Uitenbroek06} with some changes in notation, the response function can be found as 
\begin{equation}
 R_{i}(\lambda, z) = \frac{1}{\Delta B} \frac{\mathrm{d}}{\mathrm{d}z} \Delta I_i^{z}(\lambda) \, ,
 \label{eqapp::ResponsMagPert}
\end{equation}
in which the $B$ superscript is dropped from the response function for simplicity of notation. $\Delta I_i^{z}(\lambda)$ is the difference between the value of the Stokes parameter $I_i(\lambda)$ obtained with a perturbation up to height $z$ and the value obtained in the unperturbed case. 
For a discrete spatial grid, the response function for a perturbation up to grid point $k$ can be computed according to 
\begin{equation}
 R_{i}(\lambda, z_k) = \frac{1}{\Delta B} \frac{I_i^{z_k}(\lambda) - I_i^{z_{k-1}}(\lambda)}{z_{k} - z_{k-1}} \, ,
\label{eqapp::ResponsMagNum}
\end{equation}
where $z_k$ and $z_{k-1}$ are the heights at the grid point under consideration and the adjacent one closer to the lower boundary. 
${I_i^{z_k}}$ and ${I_i^{z_{k-1}}}$ are the $i$-th components of the Stokes vector, computed considering the perturbation up to grid points $k$ and $k-1$, respectively. 
The response functions for the four Stokes components over intensity are shown in Figs.~\ref{fig::RespVert} and~\ref{fig::RespHoriz} for vertical and horizontal magnetic fields, respectively.\footnote{The response functions shown in the figures were smoothed by performing a cubic interpolation onto spatial and frequency grids that are finer than those of the original problem.}   
The figures show a $0.8$~\AA--wide spectral range centered on the D${}_2$ line. 
The response functions were computed from the Stokes profiles obtained as in Sects~\ref{sec::ResBvert} and \ref{sec::ResBhor}, 
for an LOS with $\mu = 0.2$ and with $B^{(0)} = 5$~G and $\Delta B = 1$~G. 
For each Stokes component, the $R_{i}(\lambda,z_k)/I(\lambda)$ were normalized to their largest absolute value, 
and therefore they do not directly provide information on the impact of the magnetic field relative to the Stokes profiles.  
In order to highlight this impact, the corresponding Stokes profiles for each panel of Figs.~\ref{fig::RespVert} and~\ref{fig::RespHoriz} are overplotted with magenta curves, which show those obtained in the unperturbed case (solid curves) and in the case that the perturbation is included at all grid points (dashed-dotted curves). 
Indeed, the comparison of the intensity profiles that were obtained with and without such perturbations indicates that they are not appreciably impacted by magnetic fields with the strengths considered in this section, whereas the polarization patterns are clearly sensitive to such magnetic fields. 

For both horizontal and vertical magnetic fields, the strongest response of $Q/I$ is found in the core region for perturbations around $700$~km, which is well above the temperature minimum and corresponds to the lower chromosphere in the FAL-C model. A strong response in the line-core $U/I$ is found at similar heights in the case of a horizontal magnetic field, which introduces an axial asymmetry in the problem. On the other hand, 
the impact of a vertical magnetic field on the line-core $U/I$ is much more modest and is attributed mainly to the action of magneto-optical (MO) effects (e.g., Sect.~9.8 of LL04). 
The near-wing $Q/I$ and $U/I$ signals present their strongest response to perturbations at photospheric depths, although the amplitude of such signals is smaller than in the core region, and thus they are of less diagnostic interest. 
This response is attributed mainly to the action of MO effects and to the modification of the scattering polarization discussed in Sects.~\ref{sec::ResBvert} and~\ref{sec::ResBhor}, which still has an appreciable impact at these wavelengths. For the case of vertical magnetic fields, in which the response of $U/I$ is weak in the line core, the relative response is found to be stronger at near-wing wavelengths. 

As expected, the main response to the circular polarization signals is not found at the line center but instead at the wavelengths that correspond to the main $V/I$ peaks. As such, the strongest response is to perturbations centered at roughly $500$~km, which is close to the temperature minimum in the FAL-C atmospheric model, in agreement with \cite{QuinteroNoda+17}. The slight asymmetry with respect to the line center in the response of $V/I$, which is more apparent for the case of a vertical magnetic field, can be attributed to the AtO conversion mechanism (see Sect.~\ref{sec::ResBvert}). 

\begin{figure*}[!t]
\centering 
\includegraphics[width = 0.975\textwidth]{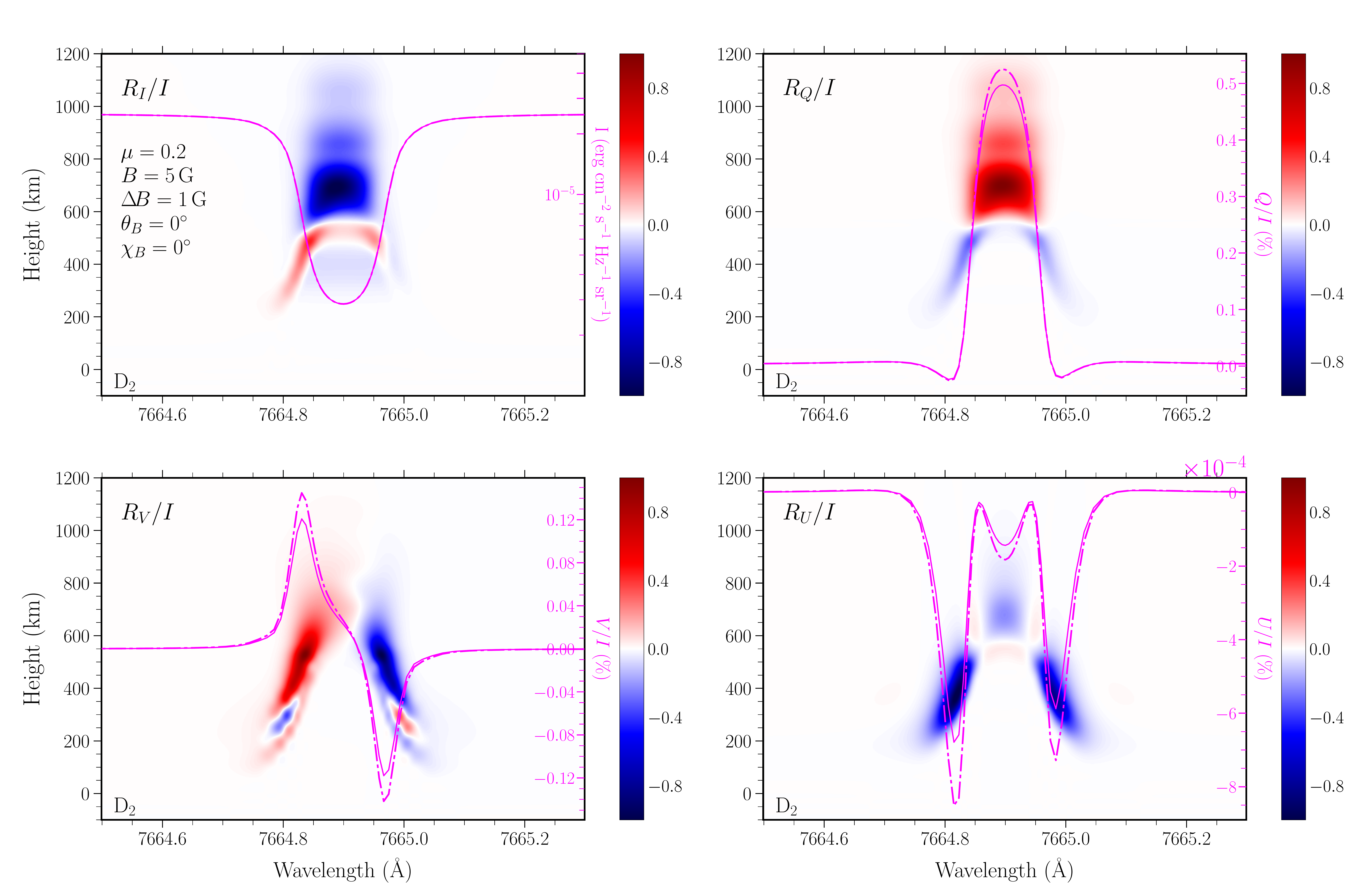}
\caption{Response functions for Stokes $I$ (upper left panel), $V$ (lower left panel), $Q$ (upper right panel), and $U$ (lower right panel)	 to a $1$~G perturbation applied to the FAL-C atmospheric model, permeated by a $5$~G vertical magnetic field (see text), over intensity. An LOS with $\mu = 0.2$ is considered. The $R_i/I$ in each panel is normalized to the largest absolute value. 
For each panel, the corresponding Stokes $I$, $V/I$, $Q/I$, or $U/I$ profiles are overplotted in magenta curves, in order to highlight the impact of the magnetic field. 
The solid curve represents the profiles obtained in the presence of the unperturbed magnetic field, whereas the dash-dotted curve represent those obtained in the case that a perturbation is applied at all height points.}
	\label{fig::RespVert} 
\end{figure*}

\begin{figure*}[!t]
\centering
\includegraphics[width = 0.975\textwidth]{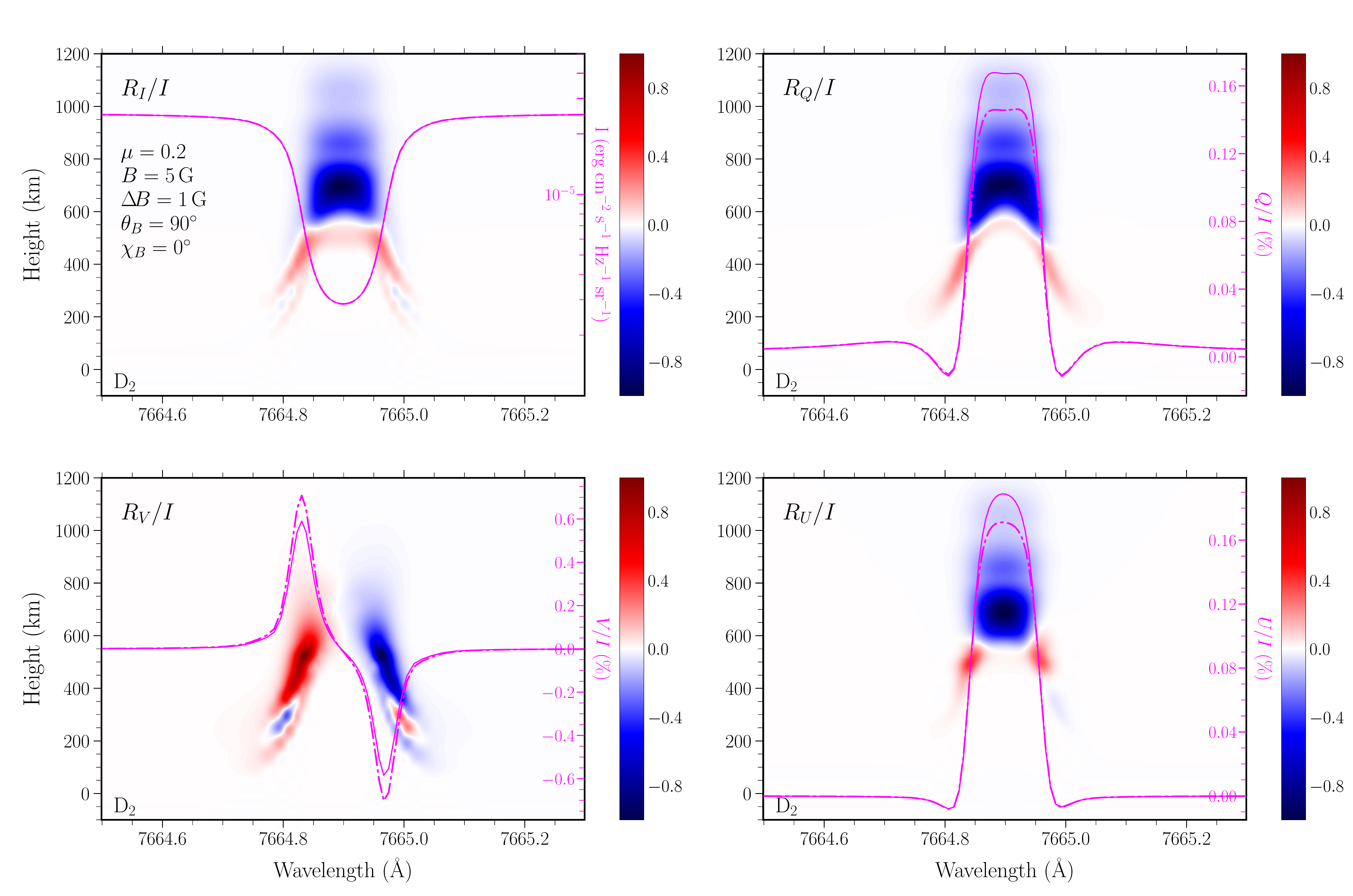}
\caption{Response functions for Stokes $I$ (upper left panel), $V$ (lower left panel), $Q$ (upper right panel), and $U$ (lower right panel) to a $1$~G perturbation applied to the FAL-C atmospheric model, permeated by a horizontal magnetic field with $\chi_B =0^\circ$, over intensity. An LOS with $\mu = 0.2$ is considered. The $R_i/I$ in each panel is normalized to the largest absolute value. 
For each panel, the corresponding Stokes $I$, $V/I$, $Q/I$, or $U/I$ profiles are overplotted in magenta curves, in order to highlight the impact of the magnetic field. 
The solid curve represents the profiles obtained in the presence of the unperturbed magnetic field, whereas the dash-dotted curve represent those obtained in the case that a perturbation is applied at all height points.}
	\label{fig::RespHoriz} 
\end{figure*}

\section{Concluding comments}
\label{sec::conclusions}
 The K~{\sc{i}} D resonance lines encode valuable information on the thermodynamical and magnetic properties of the solar regions that comprise the photosphere and the lower chromosphere. The D${}_2$ line, despite not being accessible via ground-based observations, is theoretically expected to present substantial scattering polarization signals that enhance its diagnostic value. 
In this article, a recently developed non-LTE RT code (see ABT22) was employed to determine which are the relevant physical mechanisms in shaping the Stokes profiles of the K~{\sc{i}} D lines, focusing especially on D${}_2$. 

The intensity profiles can be suitably modeled while neglecting HFS and $J$-state interference and making the CRD approximation for scattering processes. In addition, the magnetic field has no appreciable impact on the intensity of the D lines until field strengths of several hundreds of gauss are considered. The shape of these profiles depends mainly on the thermodynamic properties of the atmospheric model and their variation with height. A considerable limb darkening is also found in both D lines.  

The circular polarization signals can also be suitably modeled under the CRD approximation, without accounting for HFS or $J$-state interference. For a magnetic field with a longitudinal component of $10$~G, the resulting $V/I$ profiles present a maximum amplitude on the order of $\sim\!\!1\%$, and this amplitude increases linearly with the field strength until reaching a few hundred gauss. 
The strongest response of the $V/I$ peaks is to the magnetic fields found close to the temperature minimum. 
The AtO conversion mechanism can introduce slight asymmetries in the $V/I$ pattern of the D${}_2$ line, which are clearly appreciable in the presence of vertical magnetic fields with strengths up to $15$~G. For stronger fields, such that the upper level of D${}_2$ enters the complete 
PB effect regime, the signatures of this mechanism become less apparent. 
From a diagnostic perspective, it is noteworthy that the magnetograph formula can be suitably applied to infer the longitudinal component of weak magnetic fields from the $V/I$ patterns of both D lines, while neglecting HFS when determining the effective Land\'e factors.  

On the other hand, modeling the linear polarization profiles is significantly more involved. 
Unlike for the intensity signal, a minor but appreciable error is incurred in the scattering polarization when assuming that all potassium atoms are found in the form of the most abundant isotope, ${}^{39}\!$K. The CRD treatment of the scattering polarization signals represents only a rough approximation; it leads to an overestimation of the line-center amplitude that is still appreciable when accounting for the spectral smearing due to solar dynamics and the finite resolution of instruments such as SCIP. 
The impact of $J$-state interference can be safely neglected, but both HFS and $F$-state interference must be taken into account. 
In the absence of a magnetic field, the scattering polarization amplitude of the D${}_2$ line is found to be on the order of $0.4\%$ for an LOS with $\mu = 0.2$. This value is found when accounting for HFS, which depolarizes the signal by roughly a factor $3$. 
It is also necessary to account for HFS in order to reproduce the antisymmetric pattern found in the core region of the D${}_1$ line. However, because of the small spectral separation between the HFS components of this intrinsically unpolarizable line, its $Q/I$ peak-to-peak amplitude is on the order of $10^{-5}$, which is at the detection limit of state-of-the-art spectro-polarimeters.
A preliminary investigation of the depolarizing effect of collisions with neutral perturbers was also carried out, which indicated that they appreciably reduce the scattering polarization amplitudes of both D lines. 
A deeper analysis of this aspect is left for a forthcoming publication. 

{The response of the D${}_2$ line-core linear polarization is strongest for magnetic fields at heights that correspond to the lower chromosphere. 
The impact of the magnetic field is especially notable in the range of strengths up to $15$~G, for which its upper FS level is still in the incomplete PB effect regime for HFS. Above this range, the variation of the scattering polarization amplitude with field strength becomes more modest and it eventually reaches saturation at around $50$~G. On the other hand, in the presence of magnetic fields with a net transverse component of around $50$~G, a symmetric linear polarization pattern associated with the Zeeman effect begins to appear, and its amplitude increases for stronger magnetic fields. 
The magnetic sensitivity of the scattering polarization signals is also highly dependent on the orientation of the magnetic field, which is highlighted 
in this work through three illustrative cases. 
The presence of a vertical magnetic field induces an enhancement of the scattering polarization amplitude. By contrast, a horizontal magnetic field with $\chi_B = 0^\circ$ produces a depolarization and a rotation of the plane of linear polarization. Finally, the presence of a microstructured and isotropic magnetic field also depolarizes the $Q/I$ signal, and its amplitude at saturation is just above half of that found in the nonmagnetic case. The saturation amplitudes obtained when accounting for HFS and neglecting it coincide, in agreement with theoretical predictions. 
The CLV of the D${}_2$ scattering polarization, relative to its amplitude at the limb, is quite similar for the various orientations and for the nonmagnetic case. The only notable difference in behavior is found in the presence of horizontal fields, for which a nonzero signal is found at small inclinations, which can be attributed to the Hanle effect for forward scattering. 
These findings, together with the above-mentioned sensitivity of the circular polarization signals to the longitudinal component of the magnetic field, establish the polarization signals of the K~{\sc{i}} D lines as valuable observables for inferring information about the strength and geometry of the magnetic field in the solar regions between the upper photosphere and lower chromosphere. 

This theoretical work was focused on 1D static atmospheric models. However, recent investigations on other resonance lines, such as the Sr~{\sc{i}} line at 4607~\AA\ \citep{DelPinoAleman+18} or the Ca~{\sc{i}} line at 4227~\AA\ \citep{JaumeBestard+21} emphasized that the horizontal inhomogeneities in thermodynamical properties such as temperature and density can strongly influence their scattering polarization signals, as can the horizontal components of the macroscopic atmospheric velocities. 
Even in 1D models, it may be important to account for the dynamic nature of the solar atmosphere \citep[see the investigation by][ focused on the resonance and subordinate ultraviolet lines of Mg~{\sc{ii}}]{delPinoAleman+20}.  
Moreover, the theoretical results of \cite{QuinteroNoda+17} showed that atmospheric LOS velocities also have a considerable impact on the shape of the circular polarization signals of the K~{\sc{i}} D-lines. Thus, it will be of interest to extend the present investigation to consider dynamic and/or 3D models of the solar atmosphere. \\

\noindent{\scriptsize {\textit{Acknowledgements}. The funding received from the European Research Council (ERC) under the European Union's Horizon 2020 research and innovation programme (ERC Advanced Grant agreement No. 742265) {and from the Swiss National Science Foundation (SNSF) through Grant 200021\_175997} is gratefully acknowledged. 
Thanks are also extended to Luca Belluzzi (IRSOL) and Javier Trujillo Bueno (IAC) for an attentive reading of the manuscript and for valuable discussions.} }

\bibliographystyle{aa}


\appendix
\onecolumn

\section{The atomic model}
\label{sec::AppAtomicProps}
As discussed in Sect.~\ref{sec::formulation}, the first step of each RT calculation carried out in this article was to solve the non-LTE problem for the unpolarized case and a multilevel atomic model. The considered model included 12 levels for K~{\sc{i}} as well as the ground level of K~{\sc{ii}}. The calculations carried out using the RH code of \cite{Uitenbroek01} thus involved 12 continuum transitions and 20 line transitions, taking PRD effects into account only for the transitions to the K~{\sc{i}} D${}_1$ and D${}_2$ lines. 
For the inelastic collisions (i.e., those that induce transitions between different FS levels), only the contribution from free electrons was taken into account, and it was computed with RH following \cite{Seaton62}. 
For the calculation of the elastic collisions using RH, the van der Waals contribution from neutral hydrogen and helium atoms was considered \citep[computed following][]{Unsold55}, as well as the quadratic Stark effect contribution from free electrons and singly charged ions \citep[computed following][]{Traving60}. 

Regarding the polarized calculations of the second step of the synthesis,  a two-term atom with an unpolarized and infinitely sharp lower level was considered in the most general case, and HFS was also taken into account (see Fig.~\ref{fig::Grotrian}). 
The FS levels of the considered system are given in Table~\ref{tab::FS}, together with their corresponding energies relative to the ground state, which are given in cm$^{-1}$. It is useful to recall that the D${}_1$ (D${}_2$) line originates from the radiative transitions between the states of the ground level and the states of the upper FS level with $J = 1/2$ ($J = 3/2$). 
\begin{table*}[!h]
 \centering
 \caption{\label{tab::FS} Fine structure parameters}
 \begin{tabular}{|c|c|} \hline
  Level & Energy (cm${}^{-1}$) \\ \hline \hline 
  $4s \; {}^{2}$S${}_{1/2}$ & $0.00$ \\ \hline
  $4p \; {}^{2}$P${}_{1/2}^{\mathrm{o}}$ & $12985.185724$ \\ \hline 
  $4p \; {}^{2}$P${}_{3/2}^{\mathrm{o}}$ & $13042.89027$ \\ \hline
 \end{tabular}
\end{table*}

The RT coefficients for the considered two-term atomic model depend on the Einstein coefficient for spontaneous emission that couples the upper and lower term, $A(\beta_u L_u S \rightarrow \beta_\ell L_\ell S)$. 
Its value was determined from the Einstein coefficients for spontaneous emission of the two D lines. The following relation (e.g., Sect.~{7.8} of LL04) holds in the L-S coupling scheme 
\begin{equation*}
 A(\beta_u L_u S \rightarrow \beta_\ell L_\ell S) = \sum_{J_\ell} A(\beta_u L_u S J_u \rightarrow \beta_\ell L_\ell S J_\ell) \, .
\end{equation*}
Bearing in mind that the D lines share the same lower FS level, in the L-S scheme the Einstein coefficient should be equal for the two lines. In reality, their experimental values slightly differ, being $3.73 \cdot 10^{7}$\,s$^{-1}$ for D${}_1$ and $3.78 \cdot 10^{7}$\,s$^{-1}$ for D${}_2$ \citep{NIST_ASD}. In this paper, the value of $A(\beta_u L_u S \rightarrow \beta_\ell L_\ell S)$ was thus taken as their average.
Throughout this work, the radiative line broadening parameter $\Gamma_R$ that appears in the redistribution matrices (see ABT22) was also taken to be equal to this value. The line broadening parameters for elastic ($\Gamma_E$) and inelastic ($\Gamma_I$) collisions were taken to be equal to the rates for elastic and inelastic collisions, respectively, which were obtained in the first step of the RT calculation using RH. 

Potassium is mainly found in the form of two stable isotopes, which are $^{39}\!$K and $^{41}\!$K. They have abundances of $93.3\%$ and $6.7\%$, respectively \citep[see][]{NIST_ASD}, and they both have nuclear spin $I = 3/2$. In Table~\ref{tab::Table1}, the isotopic shifts and the HFS coefficients ${\mathcal A}_{\mbox{\scriptsize{HFS}}}$ and ${\mathcal B}_{\mbox{\scriptsize{HFS}}}$ are shown for each of the three considered FS levels and for both isotopes. 
The isotopic shifts for $^{41}\!\!$~K and the values of the magnetic dipole ${\mathcal A}_{\mbox{\scriptsize{HFS}}}$ coefficients for the upper FS levels were taken from \cite{Bendali+81}, whereas the values for the electric quadrupole ${\mathcal B}_{\mbox{\scriptsize{HFS}}}$ coefficients of the $4p \; {}^{2}$P${}_{3/2}^{\mathrm{o}}$ level were taken from \cite{Ney69}. The ${\mathcal A}_{\mbox{\scriptsize{HFS}}}$ coefficients for the ground level were taken from \cite{Arimondo+77}. 
\begin{table*}[!h]
 \centering
 \caption{Hyperfine structure parameters}
 \label{tab::Table1}
 \begin{tabular}{|c|c|c|c|} \hline
  Level & Isotopic shift (cm${}^{-1}$) & ${\mathcal A}_{\mbox{\scriptsize{HFS}}}$ (cm${}^{-1}$) & ${\mathcal B}_{\mbox{\scriptsize{HFS}}}$ (cm${}^{-1}$) \\ \hline \hline 
  \multicolumn{4}{|c|}{\tiny$^{39}\!$~K} \\ \hline 
$ 4s \; {}^{2}$S${}_{1/2}$ & $0.000$ & $7.701 \cdot 10^{-3}$ &  $0.000$\\ \hline
$ 4p \; {}^{2}$P${}_{1/2}^{\mathrm{o}}$ & $0.000$ & $9.273 \cdot 10^{-4}$ &  $0.000$\\ \hline
$ 4p \; {}^{2}$P${}_{3/2}^{\mathrm{o}}$  & $0.000$  & $2.045 \cdot 10^{-4}$ & $9.073 \cdot 10^{-5}$ \\ \hline \hline
  \multicolumn{4}{|c|}{\tiny$^{41}\!$~K} \\ \hline 
$ 4s \; {}^{2}$S${}_{1/2}$ & $0.000$ & $4.236 \cdot 10^{-3}$ & $0.000$  \\ \hline
$ 4p \; {}^{2}$P${}_{1/2}^{\mathrm{o}}$ & $7.848 \cdot 10^{-3}$ & $5.067 \cdot 10^{-4}$ & $0.000$ \\ \hline
$ 4p \; {}^{2}$P${}_{3/2}^{\mathrm{o}}$  & $7.877 \cdot 10^{-3}$& $1.134 \cdot 10^{-4}$ & $1.114 \cdot 10^{-4}$ \\ \hline 
 \end{tabular}
\end{table*}

\section{Radiative transfer coefficients with multiple isotopes}
\label{sec::AppIsotopes}
The expressions for the RT coefficients considered in this work for the second step of the RT calculation can be found in the appendices of ABT22, where one can also find the simpler expressions that correspond to approximations such as neglecting the magnetic field, neglecting $F$- or $J$-state interference, or making the assumption of CRD. In that article, however, the expressions are only given for the case in which a single isotopic species is considered. 
In the present work, a number of calculations were carried out considering the contribution from two isotopes. In the multi-isotope case, the line contribution to the RT coefficients can be given as a linear combination of the RT coefficients for each isotope, labeled with superscript $s$ and weighted by their relative abundances $n_s$, 
\begin{subequations}
\begin{align}
 & \eta^\ell_i(\nu,\mathbf{\Omega}) = \sum_s n_s \, \eta_i^{\ell\,(s)}(\nu,\mathbf{\Omega}) \, , \label{eq::etaisotope} \\
 & \rho^\ell_i(\nu,\mathbf{\Omega}) = \sum_s n_s \, \rho_i^{\ell\,(s)}(\nu,\mathbf{\Omega}) \, , \label{eq::rhoisotope} \\
 & {\varepsilon^\ell_i(\nu,\mathbf{\Omega}) = \sum_s n_s \, \varepsilon^{\ell\,(s)}_i(\nu,\mathbf{\Omega})}  \label{eq::emissotope}  \, . 
 \end{align}
  \label{eqapp::MultipleIsotope}
 \end{subequations} 
Even when the nuclear spin of two different isotopes is the same, as is the case for $^{39}\!$K and $^{41}\!$K, the energies of their various states may differ due to isotopic shifts or differences in the HFS coefficients. As a result, their $\eta_i^{\ell\,(s)}$, $\rho_i^{\ell, (s)}$, and 
$\varepsilon_i^{(s)}$ coefficients may also differ.  
As discussed in Sect.~\ref{sec::formulation}, the line emissivity can be separated into thermal and scattering contributions. For each isotope, the thermal contribution can be obtained from the corresponding $\eta_i^{\ell\,(s)}$ through a relation analogous to that of Eq.~\eqref{eq::EmisLineTherm}. Regarding the line scattering contributions $\varepsilon^{\ell,\,\mathrm{sc}\,(s)}_i$, it follows from Eq.~\eqref{eq::emissotope} that 
\begin{equation}
 \bigl[\mathcal{R}\,(\nu^\prime,\mathbf{\Omega}^\prime; \nu, \mathbf{\Omega})\bigr]_{i j} = 
 \sum_s n_s \, \bigl[\mathcal{R}^{(s)}(\nu^\prime,\mathbf{\Omega}^\prime; \nu, \mathbf{\Omega}) \bigr]_{i j} \, ,
\end{equation}
which are of course related to $\varepsilon^{\ell, \mbox{\scriptsize{sc}} \,(s)}_i$ through an analogous expression to that of Eq.~\eqref{eq::EmisLineScat}. In addition, each $\mathcal{R}^{(s)}$ can obviously be separated into corresponding ${\mathcal R}^{(s)}_{\mbox{\sc{ii}}}$ and ${\mathcal R}^{(s)}_{\mbox{\sc{iii}}}$ matrices. 

\section{Hanle diagram for the scattering polarization of the D${}_2$ line for a $90^\circ$ scattering geometry}
\label{sec::AppHanleDiag} 
\begin{figure*}[!t]
\centering
\includegraphics[width = 0.95\textwidth]{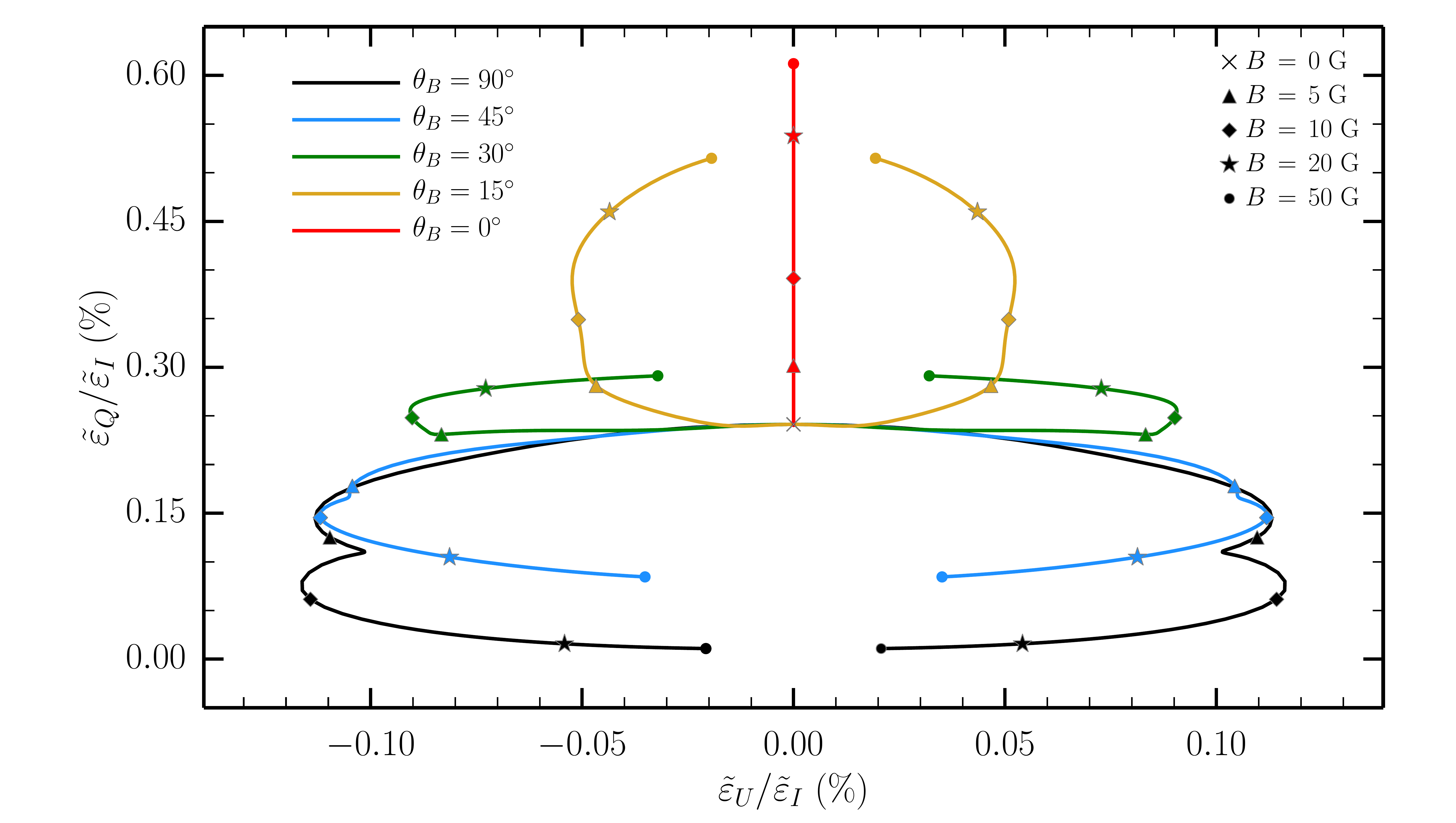}
\caption{Hanle diagram that shows the relation between the Stokes $Q$ and $U$ components of the line scattering emission vector, normalized to the intensity component, according to the strength and orientation of the magnetic field. The emission vector was obtained as described in the text and it was integrated over wavelength, taking a $3.9$~\AA--wide range centered on the D${}_2$ line. The direction of emission is taken with an inclination of $90^\circ$ with respect to the local vertical (the local vertical is parallel to the symmetry axis of the radiation field). Each of the curves shows the results of calculations that were carried out in the presence of magnetic fields with increasing strength, for various inclinations (see legend) with respect to the local vertical. The curves on the right (left) half of the diagram correspond magnetic fields with positive (negative) longitudinal components (see text). The points upon each curve correspond to field strengths of $5$, $10$, $20$, and $50$~G and are marked with triangles, diamonds, stars, and circles, respectively. The point corresponding to the calculation for $B = 0$~G, where all the curves converge, is marked with the gray cross. The reference direction for positive Stokes $Q$ is perpendicular to the local vertical.} 
	\label{fig::HanleDiag} 
\end{figure*}
The sensitivity of the scattering polarization signals of the K~{\sc{i}} D${}_2$ line to deterministic magnetic fields, which is discussed for two specific orientations in Sects.~\ref{sec::ResBvert} and~\ref{sec::ResBhor}, may be represented in a more compact manner through the Hanle diagram shown in Fig.~\ref{fig::HanleDiag}. 
Instead of solving the full non-LTE RT problem in the polarized case, the emission coefficient is considered, for an LOS with $\mu = 0.2$. It was calculated at a single spatial point, taking the thermodynamic properties and the incident radiation field that correspond to a height of $600$~km in the FAL-C atmospheric model. 
The rationale for this approach is similar to that of the so-called last scattering approximation, in which the emergent radiation is taken to be equal to the source function at the depth at which $\tau = \mu$ \citep[e.g.,][]{Sampoorna+09}. 
The incident radiation field was computed using RH, and it was axially symmetric around the local vertical. 
The diagram shows the ratios of the wavelength-integrated line scattering emission coefficients $\tilde{\varepsilon}_Q/\tilde{\varepsilon}_I$ and 
$\tilde{\varepsilon}_U/\tilde{\varepsilon}_I$, in which the ``$\ell, \mbox{sc}$'' labels have been dropped for simplicity of notation. The emission direction was selected with a $90^\circ$ inclination with respect to the symmetry axis of the incident radiation field. 
The spectral integration was carried out taking a $3.9$~\AA--wide wavelength interval whose center coincides with that of the D${}_2$ line. The figure illustrates the variation of such quantities with the magnetic field for strengths up to $50$~G (where the Zeeman effect still does not play an appreciable role in the linear polarization). Each curve in the figure represents the results of calculations in the presence magnetic fields with a different orientation. 
All considered magnetic field vectors were contained in the plane defined by the LOS and the local vertical, with inclinations $\theta_B$ of $0^\circ$, $15^\circ$, $30^\circ$, $45^\circ$, and $90^\circ$. For all inclinations except for the vertical, fields with a negative longitudinal component were taken in addition to those with a positive longitudinal component, such that they point in opposite directions. The orientation of fields with a positive component is given by the specified inclination $\theta_B$ and azimuth $\chi_B = 0^\circ$, whereas the orientation of fields with a negative longitudinal component is given by inclination $180^\circ - \theta_B$ and azimuth $\chi_B = 180^\circ$. 
The calculations were carried out considering a two-term atomic model with HFS. For numerical simplicity, only the ${}^{39}\!$K isotope is considered and the CRD approximation is used (see Sect.~\ref{sec::redis}). 

In the presence of a vertical magnetic field, the emitted radiation is polarized in the direction perpendicular to the vertical, and $\tilde{\varepsilon}_U/\tilde{\varepsilon}_I$ remains zero. 
Thus, the linear polarization amplitude, given solely by $\tilde{\varepsilon}_Q/\tilde{\varepsilon}_I$, increases monotonically with field strength, in agreement with the results shown in Fig.~\ref{fig::StokesVerticalSample}. 
Moreover, for this geometry, the scattering polarization depends on the quantum interference between states with the same quantum number $f$ and how they are impacted by the magnetic field, but not on the interference between states with different number $f$. 

When the magnetic field is inclined, the interference between states with both different and the same $f$ quantum number have an impact on the scattering polarization. For the case of horizontal magnetic fields, the characteristic small ``curls'' of level-crossing effects are found for field strengths between $5$ and $10$~G, which is fully consistent with the behavior observed in Fig.~\ref{fig::StokesHorizontalSample} (see also the theoretical discussion and the application to the Na~{\sc{i}} D${}_2$ line in Sect.~10.22 of LL04). 
As discussed above, this behavior is related to level-crossing phenomena. 
For stronger fields, the linear polarization amplitude of the scattered radiation is found to decrease monotonically, and it approaches saturation for fields with strengths around $50$~G. 
For fields with an inclination of $45^\circ$, the behavior is qualitatively similar to the one found in the presence of $90^\circ$ magnetic fields. 
For smaller inclinations, such as $15^\circ$ or $30^\circ$, the scattering polarization presents traits of the behavior found both for horizontal and for vertical fields. Similar to the behavior found for horizontal magnetic fields, the amplitude of $\tilde{\varepsilon}_U/\tilde{\varepsilon}_I$ sharply increases with the field strength while the fields are relatively weak, until a strength is reached at which this trend is reversed and its amplitude begins to decrease again. However, no curls are found for such magnetic field orientations. 
The amplitude of $\tilde{\varepsilon}_Q/\tilde{\varepsilon}_I$ is found to increase with the field strength -- similar to the behavior for a vertical magnetic field -- beyond a given value ($\sim\!\!1.5$~G for a $15^\circ$ inclination and $\sim\!\!5$~G for a $30^\circ$ inclination). 
Below such field strengths, however, $\tilde{\varepsilon}_Q/\tilde{\varepsilon}_I$ is found to slightly decrease with field strength instead. 

\section{Stokes profiles in the presence of various magnetic fields} 
\label{sec::AppExtraFig} 
This appendix contains four figures, which illustrate the sensitivity of the Stokes profiles of the K~{\sc{i}} D lines to magnetic fields that are stronger than those shown in the main text (up to $750$~G), as obtained with the abovementioned non-LTE RT code. 
Figs.~\ref{fig::D2v00s} and~\ref{fig::D1v00s} show the profiles obtained for the same geometry as in Sect.~\ref{sec::ResBvert}, but in the presence of stronger vertical magnetic fields, for the D${}_2$ and D${}_1$ lines, respectively.  
Figs.~\ref{fig::D2v900s} and~\ref{fig::D1v900s} show the profiles obtained for the same geometry as in Sect.~\ref{sec::ResBhor}, but in the presence of stronger horizontal magnetic fields (contained in the plane defined by the LOS and the local vertical), for the D${}_2$ and D${}_1$ lines, respectively. 

In the panels for $V/I$, the profiles that are obtained using the magnetograph formula (see Sect.~\ref{sec::circpol}) are also shown for the same field strengths (i.e., up to $750$~G). As expected, the relative difference between such profiles and the ones obtained through full RT calculations increases with field strength and begins to be significant at around $350$~G. The discrepancies are clearer for the case of horizontal magnetic fields, for which the longitudinal component is larger. In the case of a vertical magnetic field, it is interesting to point out that larger discrepancies between such calculations are found for the D${}_1$ line than for the D${}_2$ line (compare Figs.~\ref{fig::D2v00s} and ~\ref{fig::D1v00s}). 
Finding a definitive explanation for this behavior is beyond the scope of the present work, but it is worth recalling that the suitability of the magnetograph formula requires the magnetic field to be weak. More precisely, the magnetic splitting must be much smaller than the typical width of the line profiles (see Sect.~9.6 of LL04). It is thus noteworthy that, compared to the D${}_2$ line, the effective Land\'e factor (related to the magnetic splitting) is larger for D${}_1$, and its absorption profile in intensity is slightly narrower and appears to be more sensitive to the magnetic field. 
\begin{figure*}[!t]
\centering
\includegraphics[width = 0.975\textwidth]{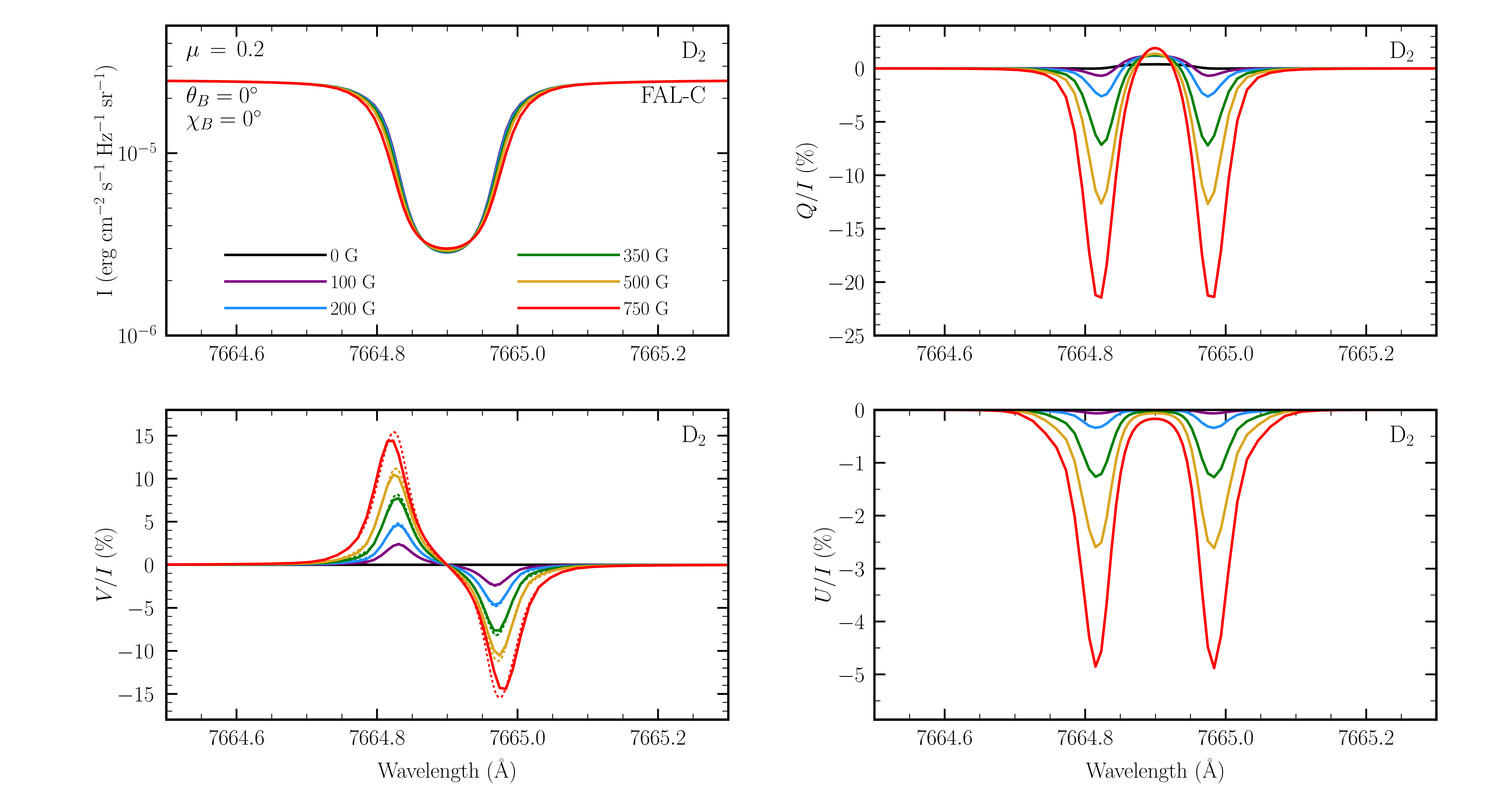}
\caption{Stokes $I$ (upper left panel), $V/I$ (lower left panel), $Q/I$ (upper right panel), and $U/I$ (lower right panel) profiles as a function of wavelength. 
The various solid colored curves represent the results of RT calculations (see main text) carried out 
in the presence of vertical ($\theta_B = 0^\circ$) magnetic fields with strengths up to $750$~G (see legend). 
The considered spectral range is centered on the K~{\sc i} D${}_2$ line. For this figure and the following ones in this appendix, the Stokes profiles are shown for an LOS with $\mu = 0.2$, resulting from calculations in which the two-term atomic model described in Sect.~\ref{sec::formulation} was considered, accounting for HFS and including only the $^{39}$K isotope. The colored dotted curves in the lower left panel represent the $V/I$ profile obtained through the magnetograph formula, for the field strengths shown in the legend.} 
	\label{fig::D2v00s} 
\end{figure*}
\begin{figure*}[!t]
\centering
\includegraphics[width = 0.975\textwidth]{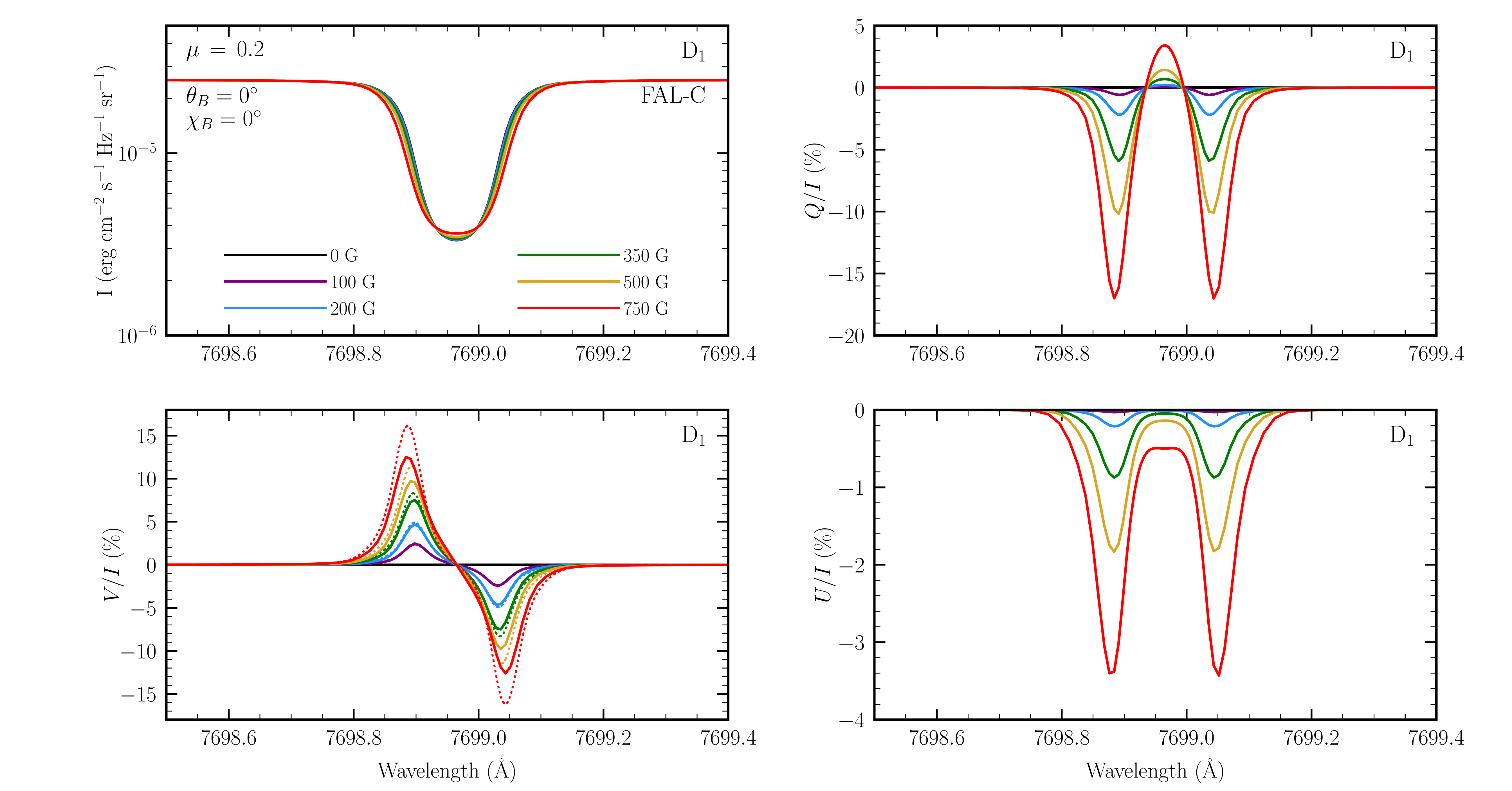}
\caption{Stokes $I$ (upper left panel), $V/I$ (lower left panel), $Q/I$ (upper right panel), and $U/I$ (lower right panel) profiles as a function of wavelength. The various solid colored curves represent the results of RT calculations (see the main text) carried out in the presence of vertical ($\theta_B = 0^\circ$) magnetic fields with strengths up to $750$~G (see legend). The considered spectral range is centered on the K~{\sc i} D${}_1$ line. The colored dotted curves in the lower left panel represent the $V/I$ profile obtained through the magnetograph formula, for the field strengths shown in the legend. }
	\label{fig::D1v00s} 
\end{figure*}

\begin{figure*}[!t]
\centering
\includegraphics[width = 0.975\textwidth]{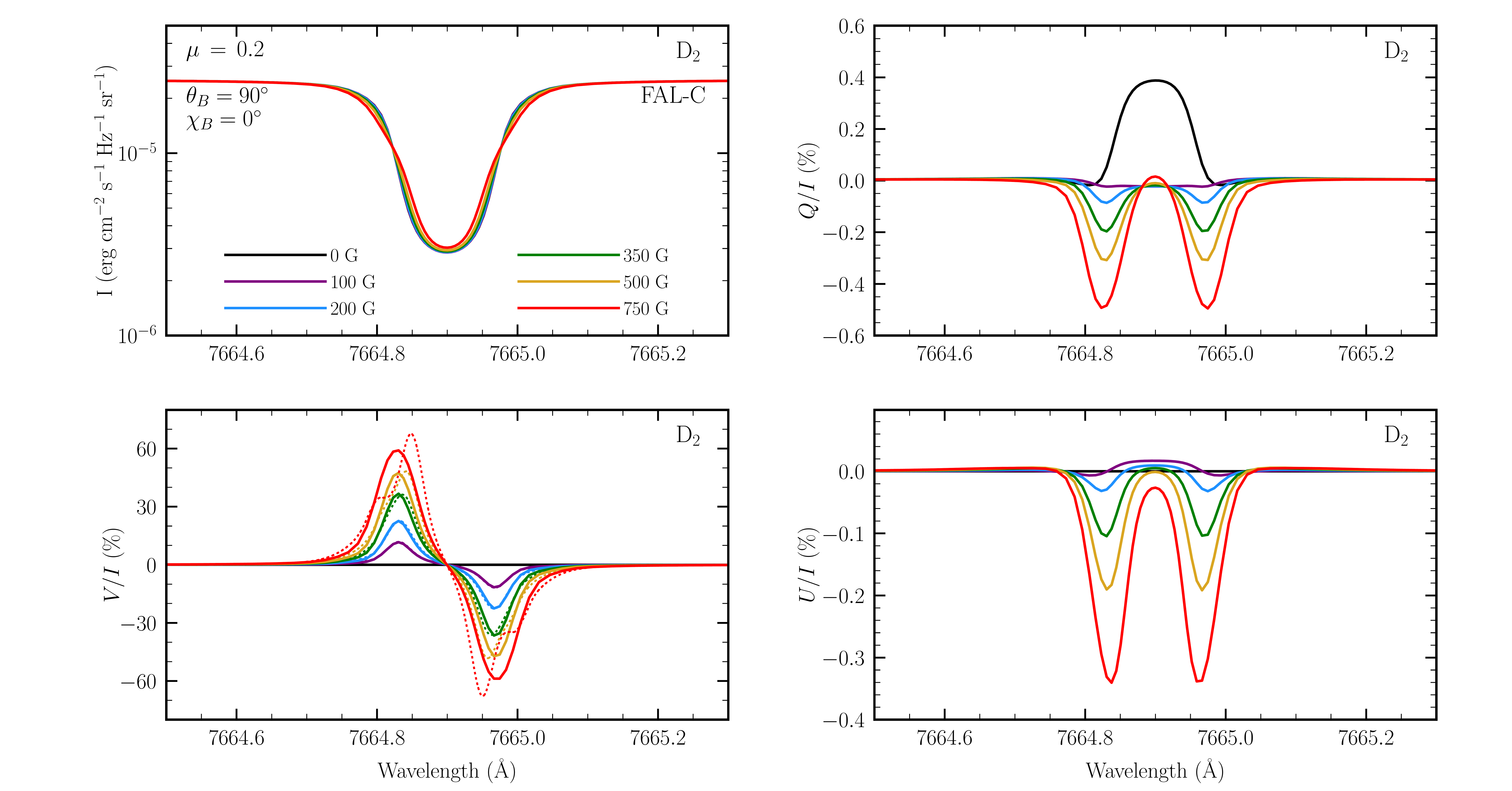}
\caption{Stokes $I$ (upper left panel), $V/I$ (lower left panel), $Q/I$ (upper right panel), and $U/I$ (lower right panel) profiles as a function of wavelength. The various solid colored curves represent the results of RT calculations (see main text) in the presence of horizontal ($\theta_B = 90^\circ$) magnetic fields with $\chi_B = 0^\circ$ and strengths up to $750$~G (see legend). The considered spectral range is centered on the K~{\sc i} D${}_2$ line. The colored dotted curves in the lower left panel represent the $V/I$ profile obtained through the magnetograph formula, for the field strengths shown in the legend.} 
	\label{fig::D2v900s} 
\end{figure*}
\begin{figure*}[!t]
\centering
\includegraphics[width = 0.975\textwidth]{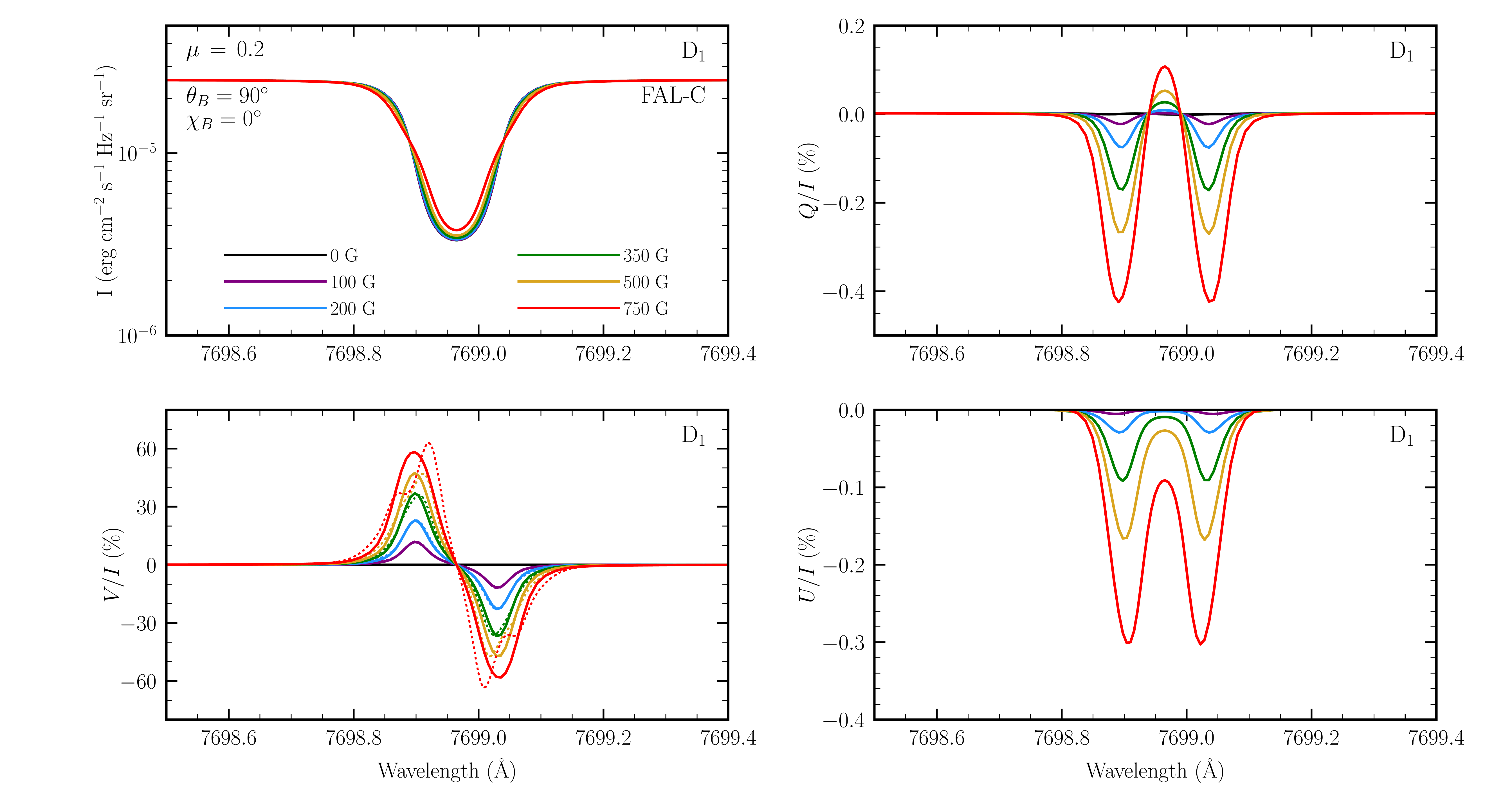}
\caption{Stokes $I$ (upper left panel), $V/I$ (lower left panel), $Q/I$ (upper right panel), and $U/I$ (lower right panel) profiles as a function of wavelength. The various solid colored curves represent the results of RT calculations (see main text) in the presence of horizontal ($\theta_B = 90^\circ$) magnetic fields with $\chi_B = 0^\circ$ and strengths up to $750$~G (see legend). The considered spectral range is centered on the K~{\sc i} D${}_1$ line. The colored dotted curves in the lower left panel represent the $V/I$ profile obtained through the magnetograph formula, for the field strengths shown in the legend.} 
	\label{fig::D1v900s} 
\end{figure*}
\end{document}